\documentclass[twocolumn,tighten,trackchanges]{aastex631}
\PassOptionsToPackage{dvipsnames,svgnames,table}{xcolor}

\usepackage{amsmath}
\usepackage{graphicx}
\usepackage[dvipsnames]{xcolor}
\usepackage{apjfonts}
\usepackage{booktabs}
\usepackage{tablefootnote}
\usepackage{lipsum}
\usepackage{xspace}

\newcommand{\kms}{\,km\,s$^{-1}$}

\newcommand{\msolar}{M$_{\odot}$}

\def\lsim{\hbox{\rlap{\raise 0.425ex\hbox{$<$}}\lower 0.65ex\hbox{$\sim$}}}
\def\gsim{\hbox{\rlap{\raise 0.425ex\hbox{$>$}}\lower 0.65ex\hbox{$\sim$}}}
\def\arcmin{\hbox{$^\prime$}}
\def\arcsec{\hbox{$^{\prime\prime}$}}
\def\arcdeg{\mbox{$^\circ$}}

\newcommand{\highzIc}{SN~2023adta\xspace}
\newcommand{\snz}{2.83}

\definecolor{maroon}{rgb}{0.760,0.118,0.337}

\newcommand{\STScI}{Space Telescope Science Institute, Baltimore, MD 21218, USA}
\newcommand{\JHU}{Physics and Astronomy Department, Johns Hopkins University, Baltimore, MD 21218, USA}

\newcommand{\NEF}{NASA Einstein Fellow}

\defcitealias{Decoursey24}{D24}

\newcommand{\ISEF}{ISEF International Fellowship}

\shorttitle{SN~Ic-BL at z=2.83}
\shortauthors{Siebert et al.}

\graphicspath{{./}{figures/}}

\begin{document}

\title{Discovery of a Relativistic Stripped Envelope Type Ic-BL Supernova at z = 2.83 with \textit{JWST}}

\correspondingauthor{Matthew~R.~Siebert}
\email{msiebert@stsci.edu}

\author[0000-0003-2445-3891]{M.~R.~Siebert}
\affiliation{\STScI}

\author[0000-0002-4781-9078]{C.~DeCoursey}
\affiliation{Steward Observatory, University of Arizona, 933 N. Cherry Avenue, Tucson, AZ 85721 USA}

\author[0000-0003-4263-2228]{D.~A.~Coulter} 
\affiliation{\STScI} 

\author[0000-0003-0209-674X]{M.~Engesser}
\affiliation{\STScI}

\author[0000-0002-2361-7201]{J.~D.\ R.\ Pierel}
\altaffiliation{\NEF}
\affiliation{\STScI}

\author[0000-0002-4410-5387]{A.~Rest} 
\affiliation{\STScI}
\affiliation{\JHU}

\author[0000-0003-1344-9475]{E.~Egami}
\affiliation{Steward Observatory, University of Arizona, 933 N. Cherry Avenue, Tucson, AZ 85721 USA}

\author[0000-0002-9301-5302]{M.~Shahbandeh} 
\affiliation{\STScI}

\author[0000-0003-1060-0723]{W.~Chen} 
\affiliation{Department of Physics,
Oklahoma State University, 145 Physical Sciences Bldg, Stillwater, OK
74078, USA}

\author[0000-0003-2238-1572]{O.~D.~Fox} 
\affiliation{\STScI}

\author[0000-0002-0632-8897]{Y.~Zenati}
\altaffiliation{\ISEF}
\affiliation{\JHU}
\affiliation{\STScI}

\author[0000-0003-1169-1954]{T.~J.~Moriya}
\affiliation{National Astronomical Observatory of Japan, National Institutes of Natural Sciences, 2-21-1 Osawa, Mitaka, Tokyo 181-8588, Japan}
\affiliation{Graduate Institute for Advanced Studies, SOKENDAI, 2-21-1 Osawa, Mitaka, Tokyo 181-8588, Japan}
\affiliation{School of Physics and Astronomy, Monash University, Clayton, Victoria 3800, Australia}

\author[0000-0002-8651-9879] {A.~J.~Bunker}
\affiliation{Department of Physics, University of Oxford, Denys Wilkinson Building, Keble Road, Oxford OX1 3RH, UK}

\author[0000-0002-1617-8917] {P.~A.~Cargile}
\affiliation{Center for Astrophysics $|$ Harvard \& Smithsonian, 60 Garden St., Cambridge MA 02138 USA}

\author[0000-0002-2678-2560] {M.~Curti}
\affiliation{European Southern Observatory, Karl-Schwarzschild-Strasse 2, 85748 Garching, Germany}

\author[0000-0002-2929-3121] {D.~J.~Eisenstein}
\affiliation{Center for Astrophysics $|$ Harvard \& Smithsonian, 60 Garden St., Cambridge MA 02138 USA}

\author[0000-0003-3703-5154]{S.~Gezari}
\affiliation{\STScI}

\author[0000-0001-6395-6702]{S.~Gomez}
\affiliation{\STScI}

\author[0000-0002-5063-0751]{M.~Guolo}
\affiliation{\JHU}

\author[0000-0002-9280-7594] {B.~D.~Johnson}
\affiliation{Center for Astrophysics $|$ Harvard \& Smithsonian, 60 Garden St., Cambridge MA 02138 USA}

\author[0000-0002-7593-8584]{B.~A.~Joshi} 
\affiliation{\JHU}

\author[0000-0003-2495-8670]{M.~Karmen}
\affiliation{\JHU}

\author[0000-0002-4985-3819] {R.~Maiolino}
\affiliation{Kavli Institute for Cosmology, University of Cambridge, Madingley Road, Cambridge CB3 0HA, UK}
\affiliation{Cavendish Laboratory, University of Cambridge, 19 JJ Thomson Avenue, Cambridge CB3 0HE, UK}
\affiliation{Department of Physics and Astronomy, University College London, Gower Street, London WC1E 6BT, UK}

\author[0000-0001-9171-5236]{R.~M.~Quimby}
\affiliation{Department of Astronomy/Mount Laguna Observatory, San Diego State University, 5500 Campanile Drive, San Diego, CA 92812-1221, USA}
\affiliation{Kavli Institute for the Physics and Mathematics of the Universe (WPI), The University of Tokyo Institutes for Advanced Study, The University of Tokyo, Kashiwa, Chiba 277-8583, Japan}

\author[0000-0002-4271-0364] {B.~Robertson}
\affiliation{Department of Astronomy and Astrophysics, University of California, Santa Cruz, 1156 High Street, Santa Cruz CA 96054, USA}

\author[0000-0002-7756-4440]{L.~G.~Strolger} 
\affiliation{\STScI}

\author[0000-0002-4622-6617]{F.~Sun}
\affiliation{Center for Astrophysics $|$ Harvard \& Smithsonian, 60 Garden St., Cambridge MA 02138 USA}

\author[0000-0001-5233-6989]{Q.~Wang} 
\affiliation{\JHU}

\author[0000-0002-4043-9400]{T.~Wevers}
\affiliation{\STScI}

\begin{abstract}
We present JWST NIRCam and NIRSpec observations of a Type Ic supernova (SN Ic) and its host galaxy (JADES-GS+53.13533-27.81457) at $z=\snz$. This SN (named \highzIc) was identified in deep James Webb Space Telescope (JWST)/NIRCam imaging from the JWST Advanced Deep Extragalactic Survey (JADES) Program.  Follow-up observations with JWST/NIRSpec provided a spectroscopic redshift of $z = 2.83$ and the classification as a SN Ic-BL. The light curve of \highzIc matches well with other stripped envelope supernovae and we find a high peak luminosity, $M_V = -19.0 \pm 0.2$ mag, based on the distribution of best-fit SNe. The broad absorption features in its spectrum are consistent with other SNe Ic-BL 1-3 weeks after peak brightness. We measure a Ca II NIR triplet expansion velocity of $29{,}000 \pm 2{,}000$ \kms. The host galaxy of \highzIc is irregular, and modeling of its spectral energy distribution (SED) indicates a metallicity  of $Z = 0.35^{+0.16}_{-0.08} Z_{\odot}$. This environment is consistent with the population of low-$z$ SNe~Ic-BL which prefer lower metallicities relative to other stripped envelope supernovae, and track long duration $\gamma$-ray burst (LGRB) environments. We do not identify any GRBs that are coincident with \highzIc. Given the rarity of SNe~Ic-BL in the local universe, the detection of a SN~Ic-BL at $z = 2.83$ could indicate that their rates are enhanced at high redshift.

\end{abstract}

\section{Introduction}\label{s:intro}

Stripped-envelope supernovae (SESNe) result from the core-collapse of massive stars that have experienced some form of mass-loss causing them to lose their outer layers of hydrogen and sometimes helium \citep{Clocchiatti+96}. A Type Ic supernova (SN~Ic) is a SESN characterized by the lack of both H and He features in its spectra \citep{Filippenko95, Matheson+01} and requires a progenitor star that has experienced a large amount of stripping \citep[see,][]{Woosley+94_IIb,Dessart+12,Gal-Yam17, Dessart+20}. Two primary channels with different mass-loss mechanisms are thought to be able to produce a SN~Ic. First, the collapse of a high-mass Wolf-Rayet (WR) star \citep{WoosleyWeaver95, Georgy09}, whose strong metal line blanketing-driven winds have removed its outer layers. Progenitor angular momentum and metallicity likely play an important role in these progenitors because they result in stronger stellar winds. Second, young massive stars in close binary systems whose mass-loss occurs through Roche-Lobe overflow or common envelope evolution \citep{Podsiadlowski+92, Yoon&Woosley10, Sana2012, Lyman16}. It is important to note that these channels are not mutually-exclusive, and a combination of both mass-loss mechanisms is likely needed to explain the diversity of SESNe \citep{Smith15}.

There have been limited searches to directly detect companion stars. To date, there are only two SNe Ic nearby enough to detect any potential companion. Deep post-explosion upper limits of the Type Ic SN 1994I, which had earlier data restricting any single star scenario, favored companion scenarios with non-conservative mass transfer with intermediate initial orbital periods and mass ratios \citep{vandyk16}. Deep, post-explosion observations of the fully stripped  Type~Ib/c SN 2013ge resulted in a direct detection of a surviving companion consistent with a slightly reddened post-main sequence 12 \msolar~star \citep{fox22}. Aside from these two post-explosion observations, pre-explosion spectral energy distributions (SEDs) and detailed modeling suggest potential companions in the Type~Ib iPTF13bvn \citep[e.g.,][]{cao13,eldridge16,folatelli16}, Ib 2019yvr \citep{kilpatrick21,sun21}, Ic 2020oi \citep{gagliano22}, although none of these have been confirmed with post-explosion imaging. 

Broad-line SNe~Ic (SNe~Ic-BL) are a rare subclass of the Type Ic events, exhibiting broad spectral features that indicate unusually high ejecta velocities of $\sim20{,}000 - 30{,}000$ \kms ($\sim0.1c$, \citealt{Galama+98Natur,Modjaz06,Modjaz+16,Sahu+18_BL,Taddia19}). Additionally, the luminosities of these SNe are typically higher than those of other core-collapse SNe \citep{Drout11, Cano13, Taddia19}. Models suggest that the kinetic energy of these explosions can be as high as 10$^{52}$ ergs \citep{Mazzali02,Maeda03,Janka16,Prentice16}, which is a factor of 10 larger than other typical SNe. Furthermore, SNe~Ic-BL are the only SN type to be associated with long-duration $\gamma$-ray bursts (LGRBs), whose gamma-ray emission lasts longer than $2$ seconds \citep{Galama+98Natur,Woosley&Bloom06, Modjaz+16, Modjaz+19}. \citet{Iwamoto99} showed these SNe can also be accompanied by $X-ray$ flashes (XRF), which further suggests that only SNe~Ic-BL are associated with LGRBs.

LGRBs are considered to be a natural result of the collapsar model for SESNe \citep{Woosley99}. In this progenitor scenario, the core of a massive star collapses to create a rapidly rotating compact object, then accretion onto this compact object utilizes the rotational energy of the star via magnetic coupling and launches collimated jets that power the explosion and produce the GRB \citep{Woosley93, Woosley99}. The absence of LGRBs and/or their associated afterglows in normal SNe Ib/c disfavors the collapsar theory for explaining these SNe, and challenges the connection between normal SNe Ic and SNe Ic-BL \citep[see,][]{Corsi+12, Barnes+18, Modjaz+19, Zenati+20}. Many SNe~Ic-BL, however, are observed without an associated LGRB, and the reason for this is not fully understood. 

Other progenitor channels for SNe Ic-BL may be possible. \citet{zapartas17} provide a detailed analysis of deep, post-explosion upper-limits of the Type Ic-BL SN 2002ap. The SN progenitor has a low metallicity and high ejecta mass, ruling out most single star models. Instead, the deep upper limits, when combined with the explosion parameters of the primary star, suggest a more likely scenario consisting of a low-mass binary system in an eccentric orbit undergoing non-conservative mass transfer. Another more exotic, but tempting, possibility for SN 2002ap is a reverse merger of the companion star with the compact remnant from the primary explosion. 

The results for SN 2002ap suggest a possible bimodality in the progenitor population of SNe Ic-BL, but one important avenue for understanding the progenitors of SESNe and the connection between SNe~Ic-BL and LGRBs is through detailed analysis of their environments. Theory predicts that metallicity should play an important role in the production of LGRBs. Specifically, \citet{Woosley06} suggested that angular momentum loss due to the metal line-driven winds in a WR star could prohibit the formation of a compact object that was rotating fast enough to launch a jet. Therefore, they proposed a metallicity threshold of $Z < 0.3 Z_{\odot}$ for collapsars. 

Several studies have confirmed that SNe~Ic-BL prefer lower-metallicity environments relative to other core-collapse SNe \citep{Sanders12, Kelly12, Arcavi18, Modjaz20}. In particular, \citet{Modjaz20} found that SNe~Ic-BL with and without associated GRBs prefer statistically similar environments with low-metallicity and high-specific star formation rates (sSFRs).  \citet{Modjaz20} conclude that SNe~Ic-BL without GRBs in their sample either produced jets that were choked within the star \citep{Milisavljevic15, Modjaz+16}, or produced off-axis GRBs. Nonetheless, it is still debated whether metallicity is required to explain the presence of jet \citep{Mannucci11}. 

The clear connection between SESNe type and their host environment suggests that metallicity could be a key factor in discerning their progenitor scenarios. Observing SESNe at high-$z$ presents an opportunity to study their properties in a diversity of environments. At high-$z$ there is more access to very low metallicity environments ($Z < 0.1 Z_{\odot}$) like those present in local metal-poor dwarf galaxies. So far only a small sample of low-$z$ core collapse SNe have been studied in extreme environments like these \citep{Arcavi10,Taddia16,Anderson18,Gutierrez18,Tucker24}. Given the rarity of SNe~Ic-BL, we do not yet know if a population of these SNe exists in very-low metallicity environments. If the environments of SNe~Ic-BL deviate from the LGRB population at high-$z$, it may indicate that multiple progenitor scenarios are required to explain these SNe. Furthermore, we can constrain how their observational properties and rates may evolve with redshift. 

In this work, we present \textit{James Webb Space Telescope} (\textit{JWST}) NIRCam and NIRSpec observations of \highzIc. This SN was identified in the JWST Advanced Deep Extragalactic Survey (JADES, \citealt{Eisenstein23}). A \textit{JWST} Director's Discretionary Time (DDT) program was accepted to follow-up the most interesting transients ($1 < z < 4$) in this field with additional NIRCam photometry and NIRSpec spectroscopy (\citealt{Decoursey24}, hereafter \citetalias{Decoursey24}). Details on the sample of high-$z$ SNe and the design of the JADES Transient Survey + DDT observations are presented in this companion paper. 

In combination with the resulting light curve from these data, we used the NIRSpec Prism spectrum of \highzIc (covering $0.6\mu$m $-$ $5.3\mu$m at $R\sim 100$) to confirm its redshift ($z=\snz$), and classify it as a SN~Ic-BL. To date, this is the highest-redshift SESN ever discovered. We also determine that it came from a low-metallicity environment which is consistent with low-$z$ analogs.

In \autoref{s:obs}, we summarize our observations, in \autoref{s:anal} we describe the observational properties of \highzIc (and its host galaxy), we discuss its classification as a SN~Ic-BL, and put it into the context of other local SESNe, and in \autoref{s:conc} we summarize our findings, their implications, and discuss future observations. In all subsequent analysis we assume a standard $\Lambda$CDM cosmology with $H_0=70$km s$^{-1}$ Mpc$^{-1}$, $\Omega_m=0.315$.




\section{Observations \& Data Reduction}\label{s:obs}

A detailed description of the JADES transient observing program methods and data reduction are presented in \citetalias{Decoursey24}. In short, JADES observations were taken acquired over a 1 year time baseline from September 2022 to January 2024, allowing us to conduct a transient search of unprecedented $5\sigma$ m$_{AB}\sim30$ depth. The first observing window $2022$ September $29$-October $5$, and the second epoch took place between $2023$ September $29$-October $3$ with an overlap of $25\arcmin^2$ in the NIRCam F090W, F115W, F150W, F200W, F277W, F335M, F356W, F410M, F444W filters. Additional visits on $2023$ November $15$ and $2024$ January $1$ were performed due to failed observations. 

A \textit{JWST} DDT program was approved to follow and classify the most interesting transients with two additional NIRCam visits on $2023$ November $28$ and $2024$ January $1$. The second visit including the NIRSpec multi-object spectroscopy (MOS) mode using the micro-shutter assembly (MSA) and Prism (R$\sim100$) grating. The MSA provided SN spectra for $\sim10$ transients, some of which are described in companion papers (e.g., \citealt{Pierel24}, Coulter et al., Egami et al. in preparation), as well as a variety of galaxy spectra. In this section, we describe the data products and reduction methods for \highzIc.

\subsection{\textit{JWST} NIRCam} \label{s:obs_phot}

\begin{figure*}[htb!]
    \centering
    \includegraphics[width=3.2in]{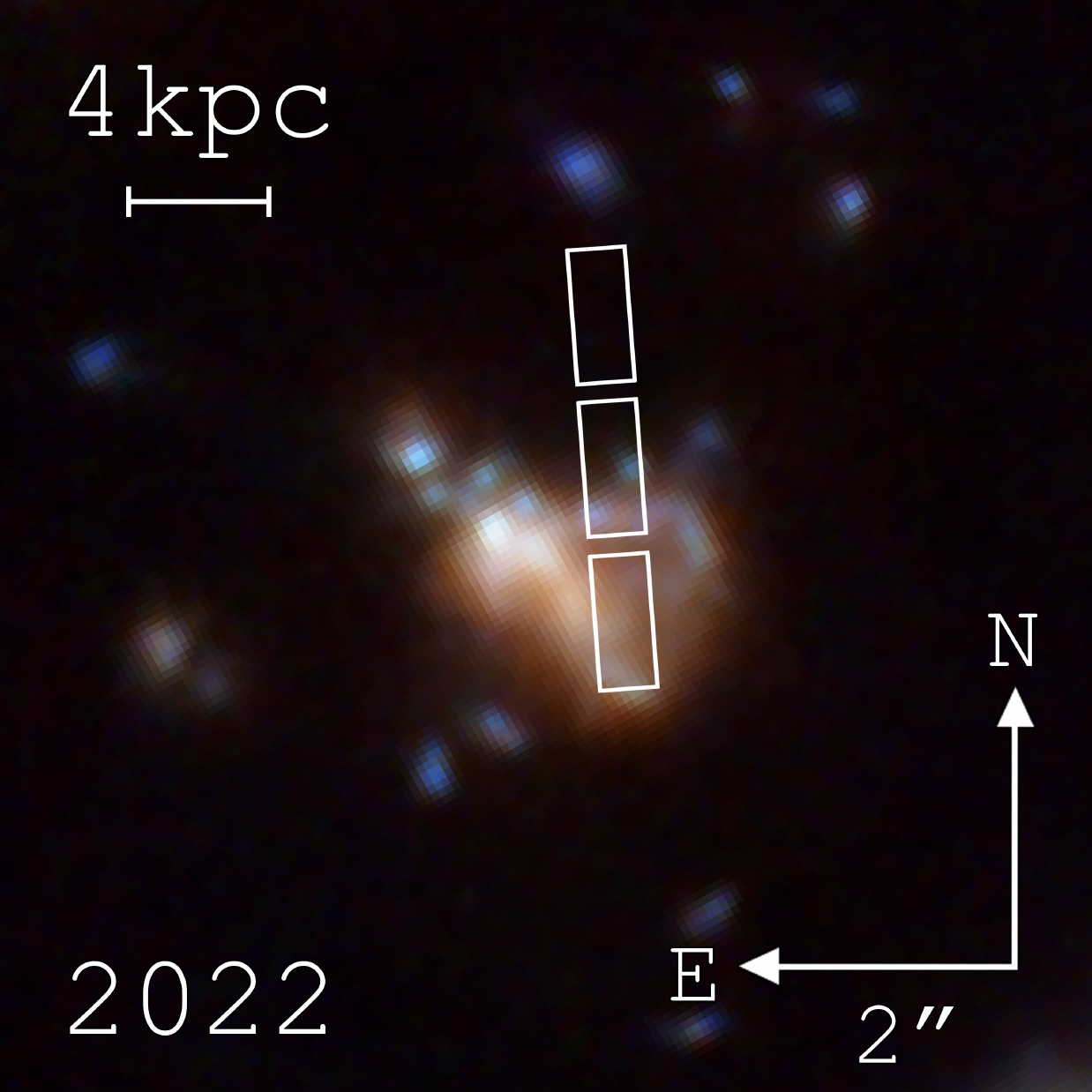}
    \includegraphics[width=3.2in]{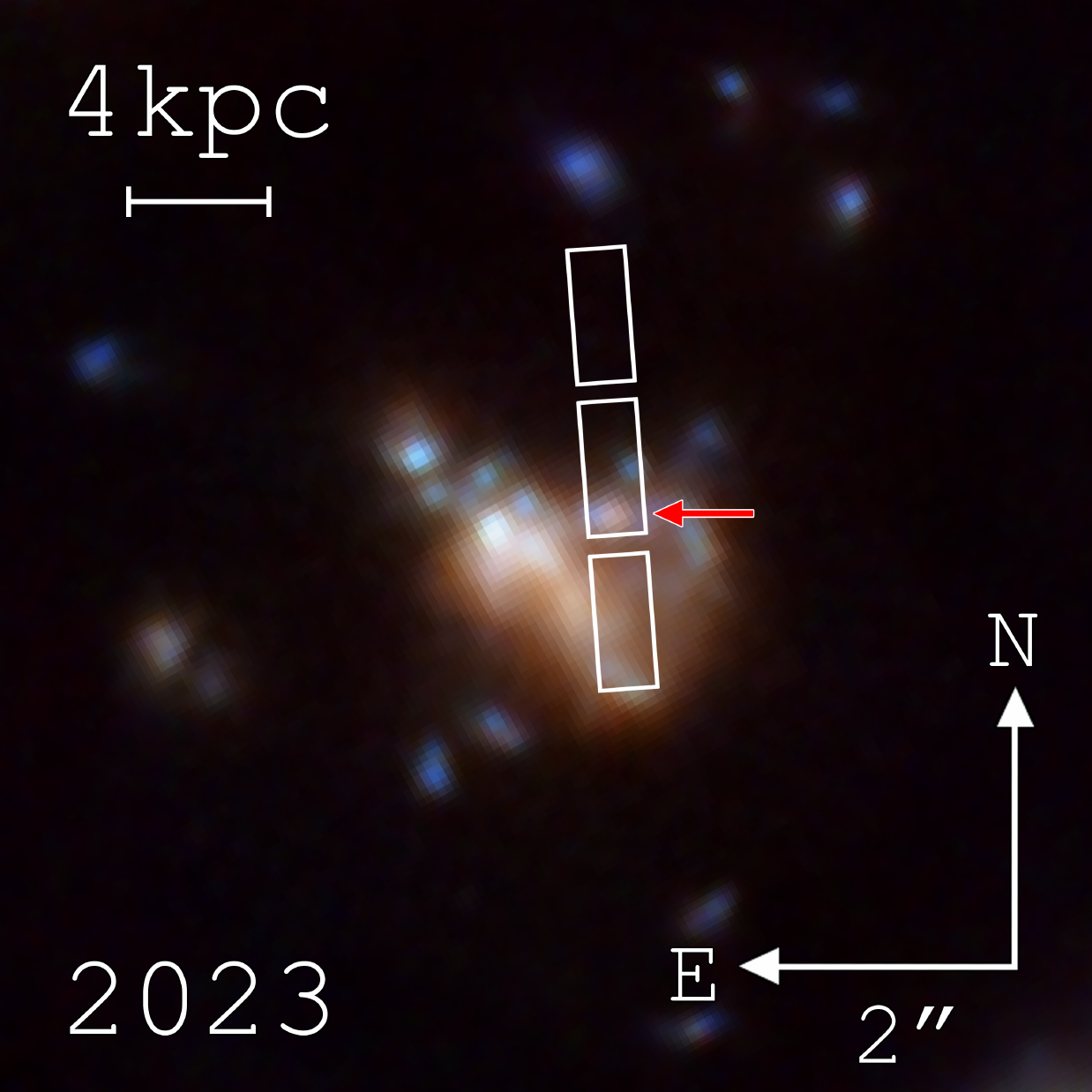}
    \caption{False color images of \highzIc and its host galaxy taken in 2022 (left) and 2023 (right). The white rectangles show the orientation of the MSA 3-shutter slitlet on sky. Two additional nodded exposures were acquired north and south of this central location. The SN (indicated by the red arrow) is located near the southern edge of the central slit.
    \label{fig:diff}}
\end{figure*}

Our data reduction methods are described in detail by \citetalias{Decoursey24}, but we summarize the process here. We adopt the point-spread function (PSF) fitting method developed in \citet{pierel_jwst_2024} for measuring photometry on Level 3 (drizzled and resampled 2D data) \textit{JWST} images. Unlike their scenario though, we have a template image for all epochs of \highzIc from the $2022$ JADES observations. We therefore first align the Level 2 (CAL) NIRCam images containing \highzIc to the template images (in each filter) using the \textit{JWST}/\textit{HST} Alignment Tool \citep[{\tt JHAT};][]{rest_arminrestjhat_2023}\footnote{\url{https://jhat.readthedocs.io}} software and then produce aligned Level 3 images with the \textit{JWST} pipeline \citep[v$1.12.5$;][and see Figure \ref{fig:diff}]{Bushouse_JWST_Calibration_Pipeline_2024}. We obtain difference images in all filters using the High Order Transform of PSF and Template Subtraction \citep[{\tt HOTPANTS};][]{becker_hotpants_2015}\footnote{\url{https://github.com/acbecker/hotpants}} code, and  then implement the Level 3 PSF fitting routine from \citet{pierel_jwst_2024} using $5\times5$ pixel cutouts and Level 2 PSF models from {\tt webbpsf}\footnote{\url{https://webbpsf.readthedocs.io}}, which are temporally and spatially dependent and include a correction to the infinite aperture flux. The Level 2 PSF models are drizzled to create a Level 3 PSF model consistent with the observations. The total measured fluxes, which are in units of MJy/sr, are converted to AB magnitudes using the native pixel scale of each image ($0.03\arcsec/$pix for SW, $0.06\arcsec/$pix for LW). Measured photometry is given in \autoref{tab:phot}.


\begin{table}
    \centering
    \caption{\label{tab:phot} Photometry for \highzIc measured in Section \ref{s:obs_phot}. }
    
    \begin{tabular*}{\linewidth}{@{\extracolsep{\stretch{1}}}*{5}{c}}
\toprule
PID&Instrument&MJD&\multicolumn{1}{c}{Filter/Disperser}&\multicolumn{1}{c}{m$_{AB}$}\\
\hline
$1180$&NIRCam&$60220$&F090W&$>30.2$\\
$1180$&NIRCam&$60220$&F115W&$29.65\pm0.11$\\
$1180$&NIRCam&$60220$&F150W&$29.11\pm0.09$\\
$1180$&NIRCam&$60220$&F200W&$28.54\pm0.08$\\
$1180$&NIRCam&$60220$&F277W&$28.91\pm0.10$\\
$1180$&NIRCam&$60220$&F335M&$28.60\pm0.13$\\
$1180$&NIRCam&$60220$&F356W&$28.69\pm0.10$\\
$1180$&NIRCam&$60220$&F410M&$29.19\pm0.22$\\
$1180$&NIRCam&$60220$&F444W&$29.31\pm0.22$\\
\hline
$1180$&NIRCam&$60264$&F090W&$>29.9$\\
$1180$&NIRCam&$60264$&F115W&$29.73\pm0.19$\\
$1180$&NIRCam&$60264$&F150W&$27.45\pm0.06$\\
$1180$&NIRCam&$60264$&F200W&$26.53\pm0.04$\\
$1180$&NIRCam&$60264$&F277W&$26.64\pm0.04$\\
$1180$&NIRCam&$60264$&F335M&$26.52\pm0.04$\\
$1180$&NIRCam&$60264$&F356W&$26.74\pm0.04$\\
$1180$&NIRCam&$60264$&F410M&$26.86\pm0.06$\\
$1180$&NIRCam&$60264$&F444W&$27.07\pm0.06$\\
\hline
$6541$&NIRCam&$60276$&F115W&$>28.9$\\
$6541$&NIRCam&$60276$&F150W&$27.63\pm0.09$\\
$6541$&NIRCam&$60276$&F200W&$26.55\pm0.05$\\
$6541$&NIRCam&$60276$&F277W&$26.57\pm0.05$\\
$6541$&NIRCam&$60276$&F356W&$26.59\pm0.06$\\
$6541$&NIRCam&$60276$&F444W&$26.92\pm0.08$\\
\hline
$6541$&NIRCam&$60310$&F150W&$28.44\pm0.14$\\
$6541$&NIRCam&$60310$&F200W&$26.98\pm0.05$\\
$6541$&NIRCam&$60310$&F277W&$26.88\pm0.07$\\
$6541$&NIRCam&$60310$&F356W&$26.69\pm0.06$\\
$6541$&NIRCam&$60310$&F444W&$26.94\pm0.09$\\
$6541$&NIRSpec&$60310$&Prism&--\\
\hline
$1180$&NIRCam&$60311$&F090W&$>29.4$\\
$1180$&NIRCam&$60311$&F115W&$>29.9$\\
$1180$&NIRCam&$60311$&F150W&$28.85\pm0.10$\\
$1180$&NIRCam&$60311$&F200W&$27.01\pm0.04$\\
$1180$&NIRCam&$60311$&F277W&$26.79\pm0.04$\\
$1180$&NIRCam&$60311$&F335M&$26.42\pm0.04$\\
$1180$&NIRCam&$60311$&F356W&$26.63\pm0.04$\\
$1180$&NIRCam&$60311$&F410M&$26.57\pm0.05$\\
$1180$&NIRCam&$60311$&F444W&$26.79\pm0.05$\\
\hline
\hline
    \end{tabular*}
\begin{flushleft}
\tablecomments{Upper limits are 5$\sigma$.}
\end{flushleft}
\vspace{1cm}
\end{table}


\subsection{\textit{JWST} Spectroscopy} 
\begin{figure*}[htb!]
    \centering
    \includegraphics[width=6.85in]{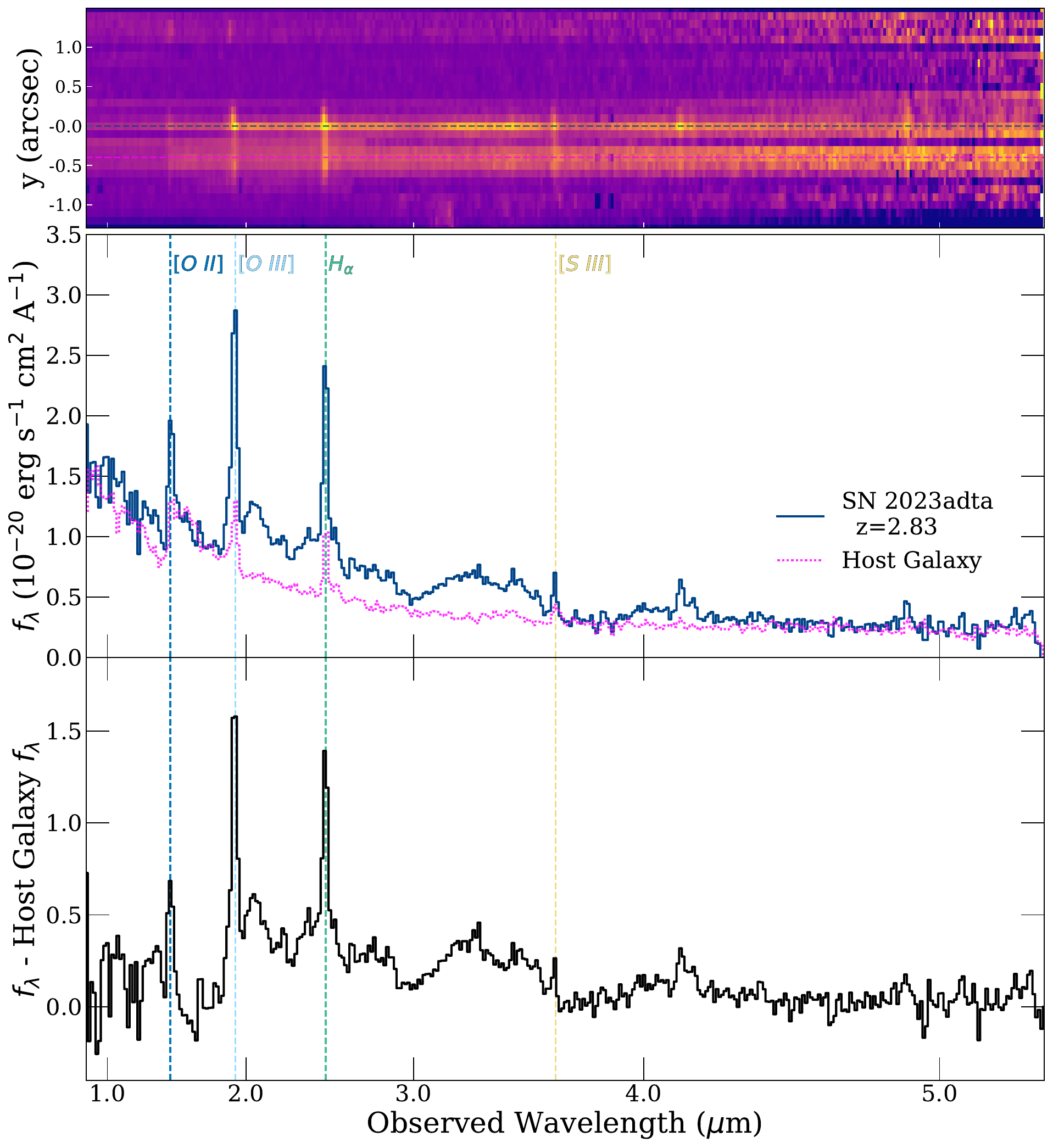}
    \caption{(\textit{top}): The 2D, master-background subtracted, NIRSpec Prism spectrum of \highzIc and its host galaxy in MJy/steradian. As shown in \autoref{fig:diff}, our observations capture both the SN and part of its host galaxy. The central rows for the extraction of \highzIc and its host galaxy are noted by the dashed-blue and pink horizontal lines, respectively. (\textit{middle}): The resulting 1D extractions of \highzIc (blue) and its host galaxy (pink) in flux converted to $f_{\lambda}$. The spectrum of \highzIc was scaled to its photometry at the same epoch, and the spectrum of the host galaxy was scaled to the 2022 photometry at the position of the SN with an identical $0.1''$ aperture. Notable narrow emission lines from the host galaxy are marked with the dashed-vertical lines. (\textit{bottom}): The difference between the flux-calibrated SN and host galaxy spectra from the middle panel. Assuming the host galaxy emission underlying the SN is similar to the galaxy emission at our extracted location, this spectrum should be mostly isolated SN flux. However, the presence of narrow emission lines could indicate that this subtraction is imperfect.
    \label{fig:2d1d}}
\end{figure*}


\begin{table}[h!]
    \centering
    \caption{JWST \highzIc NIRSpec Observation Details\label{tab:details}}
    \begin{tabular}{ll}
        \hline
        Instrument & NIRSpec  \\
        Mode & MOS  \\
        Wavelength Range & 0.6 $-$ 5.3$\mu$m  \\
        Slit & 3 Shutter ($0.46'' \times 0.2''$ each)  \\
        Grating/Filter & Prism/CLEAR  \\
        R $=\lambda/\Delta\lambda$ & $\sim 30-300$ \\
        Readout Pattern & NRSIRS2  \\
        Groups per Integration & 19  \\
        Integrations per Exposure & 2  \\
        Exposures/Nods & 3  \\
        Total Exposure Time & 16,631s
    \end{tabular}
\end{table}

\highzIc (R.A. $=3$h$32$m$32.4657$s decl. $=-27$d$48$m$52.2371$s) was selected as one of the highest-priority targets for spectroscopic follow-up observations because of its brightening in Nov 28, 2023 images (F200W $= 26.5$ mag) relatively to the discovery image taken on Oct 1, 2022 (F200W $= 28.5$ mag). We observed \highzIc on 2024 January 1 with the JWST micro-shutter assembly (MSA, \citealt{Ferruit22}) and NIRSpec Prism. Three nodded exposures were acquired using 3-shutter slitlets centered on each target in the MSA design. This observing pattern results in five unique shutter locations (each 0.46''$\times$ 0.20'') open to sky around each target. We show the orientation of the central 3-shutter slitlet for \highzIc in \autoref{fig:diff}. This orientation allowed for the independent extractions of \highzIc and part of its extended host galaxy. 

We reduced the \textit{JWST} data using the ``jwst''\footnote{\url{https://github.com/spacetelescope/jwst}} pipeline \citep[version 1.14.0;][]{Bushouse_JWST_Calibration_Pipeline_2024} routines for bias and dark subtraction, background subtraction, flat-field correction, wavelength calibration, flux calibration, rectification, outlier detection, and resampling. Given that the host galaxy is extended and its flux falls within multiple shutters, the default nodded point source background subtraction method is not sufficient for this target. Therefore we re-ran Stage 2 of the pipeline (calwebb\_spec2) using the master background subtraction strategy which makes use of designated background shutters. We performed manual interactive extractions of both \highzIc and its host galaxy from the 2D Stage 2 data product using the Specviz module of Jdaviz\footnote{\url{https://jdaviz.readthedocs.io/en/latest/specviz/index.html}}. Since the resolution of the Prism is non-linear with wavelength ($R\sim 30 - 300$, \citealt{Jakobsen2022}), we use an irregularly spaced wavelength grid with $\Delta \lambda$ ranging $34 - 201$~\AA. 

Some details of our data reduction choices and results are shown in \autoref{fig:2d1d}. In the top panel, we show the 2D, master-background subtracted NIRspec Prism spectrum of \highzIc and its host galaxy. The colormap has units of MJy/steradian. The trace of \highzIc is shown by the dashed-blue horizontal line, and the trace of the host galaxy trace is indicated by the dashed-pink horizontal line and is located 0.4'' south of \highzIc. Using the Specviz module of JDaviz \citep{Jdaviz}, we perform a boxcar extraction with 3-pixel width on both of these traces. 

We flux calibrate the host galaxy spectrum in two different ways. First, we scale the the spectrum to the 2022 host galaxy photometry\footnote{JADES Host ID 198373 from \url{https://archive.stsci.edu/hlsp/jades}} \citep[well before the SN explosion;][]{Eisenstein23}. We use this spectrum as input for the SED fitting code {\tt BAGPIPES}\footnote{\url{https://bagpipes.readthedocs.io/en/latest/}} \citep{Carnall18, Carnall19} to derive an initial redshift, and the host galaxy parameters presented in \autoref{s:host}. Second, we scale the original extracted host spectrum to the 2022 host galaxy photometry measured at the SN position in order to get an estimate of the underlying host galaxy flux. We do similarly for the extracted spectrum of \highzIc but instead scale to the NIRCam photometry observed at the same epoch. The flux-calibrated spectra of \highzIc and its host-galaxy (SN-position estimate) are shown in the middle panel of \autoref{fig:2d1d} (blue and pink curves, respectively), and the difference between them is shown in the bottom panel (black). 

Clear narrow host-galaxy emission from [\ion{O}{2}], [\ion{O}{3}], H$\alpha$, and [\ion{S}{3}], are present in the spectrum of \highzIc (marked with dashed-vertical lines). The host galaxy exhibits a strong Balmer break (1.5$\mu$m observer frame) and has notably weaker emission lines relative to the continuum than what is present in the spectrum of \highzIc. We expect that our measured host galaxy continuum flux (determined from the extended emission in the southern-most shutter) should be a reasonable estimate for what is underlying the SN. However, we note that the SN is $0.1$'' from a region of star formation whose flux also contributes to our SN spectrum (see \autoref{fig:diff}). This could be the reason for the imperfect subtraction of the host galaxy emission lines. 

The difference spectrum should contain mostly isolated SN flux. In this final spectrum of \highzIc, we see broad absorption features at $3.0\mu$m and $3.7\mu$m. We analyze these features in more detail in \autoref{s:spec_comp}.

\subsection{Redshift and Host Galaxy} \label{s:red}

\highzIc was identified in host galaxy JADES-GS+53.13533-27.81457. We fit the NIRspec Prism data of the host galaxy using the SED fitting code {\tt BAGPIPES} \citep{Carnall18, Carnall19} to determine an initial redshift estimate ($z=\snz$). We then further refine this estimate by fitting individual emission lines present in the spectrum following the methods of \citet{Bunker23}. At this redshift, several galaxy emission lines blended together. We model H$\alpha + $\ion{N}{2}, and [\ion{S}{2}] $\lambda\lambda$ $6716,6731$ as individual Gaussian components. We fit H$\alpha +$\ion{N}{2} and [\ion{S}{2}] $\lambda\lambda$ $6716,6731$, simultaneously with their centroids fixed relative to one another, and do similarly for [\ion{O}{3}] $\lambda\lambda$ $4959,5007$. Using these gas-phase nebular emission lines we measure a final redshift of $z=2.830 \pm 0.001$. This uncertainty is consistent with those measured for Prism spectra in \citet{Bunker23} ($\Delta z = 0.001 - 0.01$ at $z=\snz$). For a detailed analysis of the properties of the host galaxy of \highzIc see \autoref{s:host}

\section{Analysis}\label{s:anal}

\subsection{Light Curve}\label{s:lc}

\begin{deluxetable*}{ccccc}
    \tablecaption{Best-fit light curve parameters and properties for \highzIc \label{tab:lcfit}}
    \tablehead{\colhead{Parameter} & \colhead{Bounds} & \colhead{SN~Ib/c} & \colhead{SN~II} & \colhead{SN~Ia}}
    \tablecolumns{5}
        \startdata
        $z$&Fixed&$z=\snz$ & $z=\snz$& $z=\snz$\\
        $t_{pk}$&[60120,60411]& $60270.09\pm 0.32$ & $60255.75\pm 0.51$ & $60278.45\pm 0.41$\\
        $A$&[0,$6\times10^{-18}$]& $(4.86\pm .05)\times10^{-19}$ & $(2.73\pm .03)\times10^{-19}$& ...\\
        $M_V$&--& $19.0\pm0.2$$^{\dagger}$ & $19.4\pm0.3$$^{\dagger}$ & $19.03\pm0.03$\\
        Host $E(B-V)$&[0,3]& $(1.17\pm 4)\times10^{-3}$ & $0.43\pm 0.0006$ & ...\\
        Host $R_{V}$ &[2,4]& $2.92\pm 0.08$ & $2.03\pm 0.05$ & ...\\
        $x_0$&[0,$3\times10^{-6}$]& ... & ... & $(5.92\pm 0.05)\times10^{-8}$\\
        $x_1$&[-3,3]& ... & ... & $-0.29\pm 0.08$\\
        $c$&[-1.5,1.5]& ... & ... & $0.63\pm 0.02$ \\
        $\chi^{2}/\nu$&...& 1.00 ($1.6\pm0.4$)$^{\dagger}$ & 1.56 ($2.4\pm0.5$)$^{\dagger}$ & 1.75
        \enddata
        \tablenotetext{\dagger}{Mean and standard deviation for the 10 best-fit SNe of this type.}
\end{deluxetable*}

\begin{figure*}[htb!]
    \centering
    \includegraphics[width=2.0in]{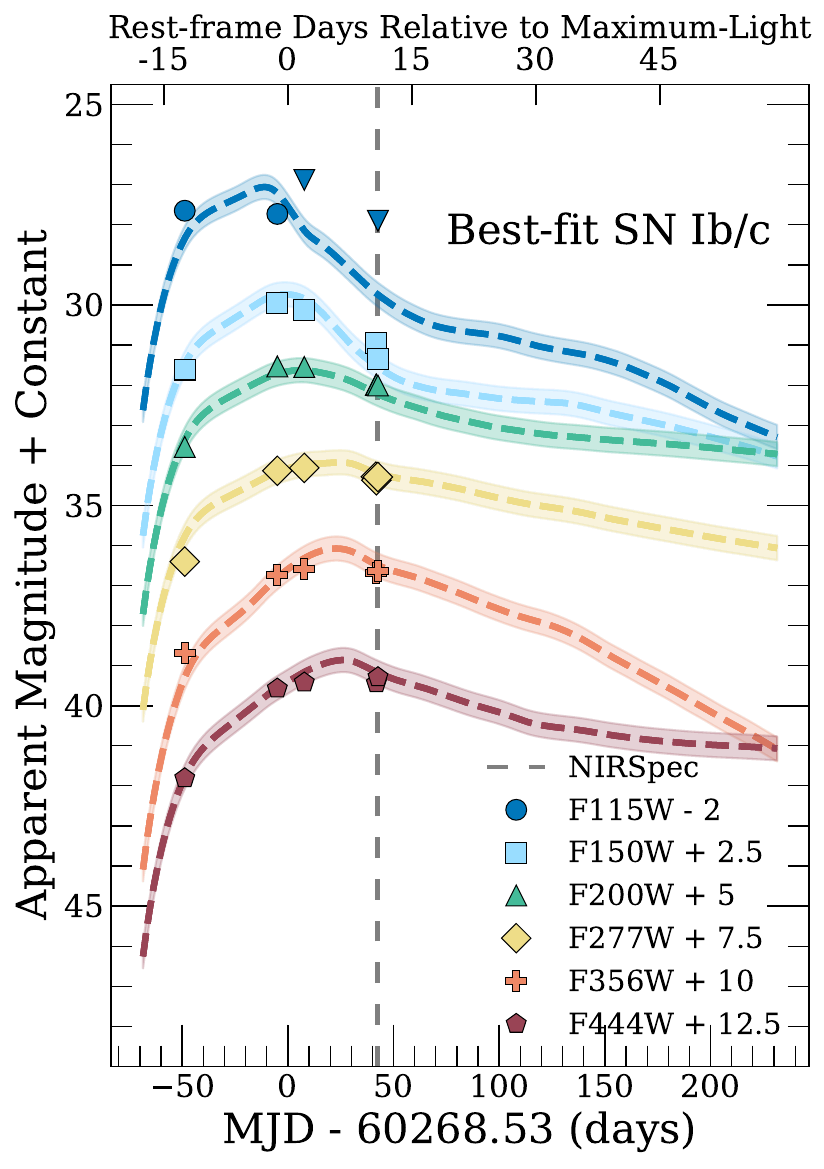}
    \includegraphics[width=2.0in]{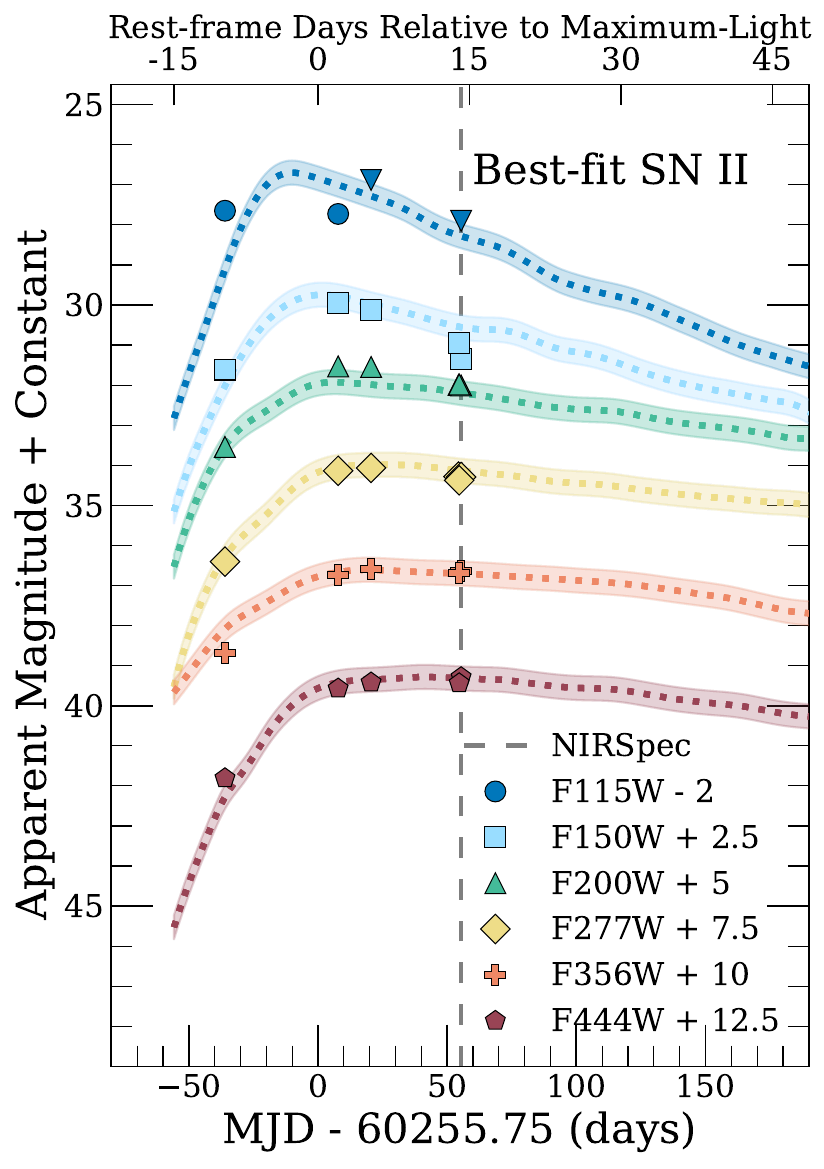}
    \includegraphics[width=2.0in]{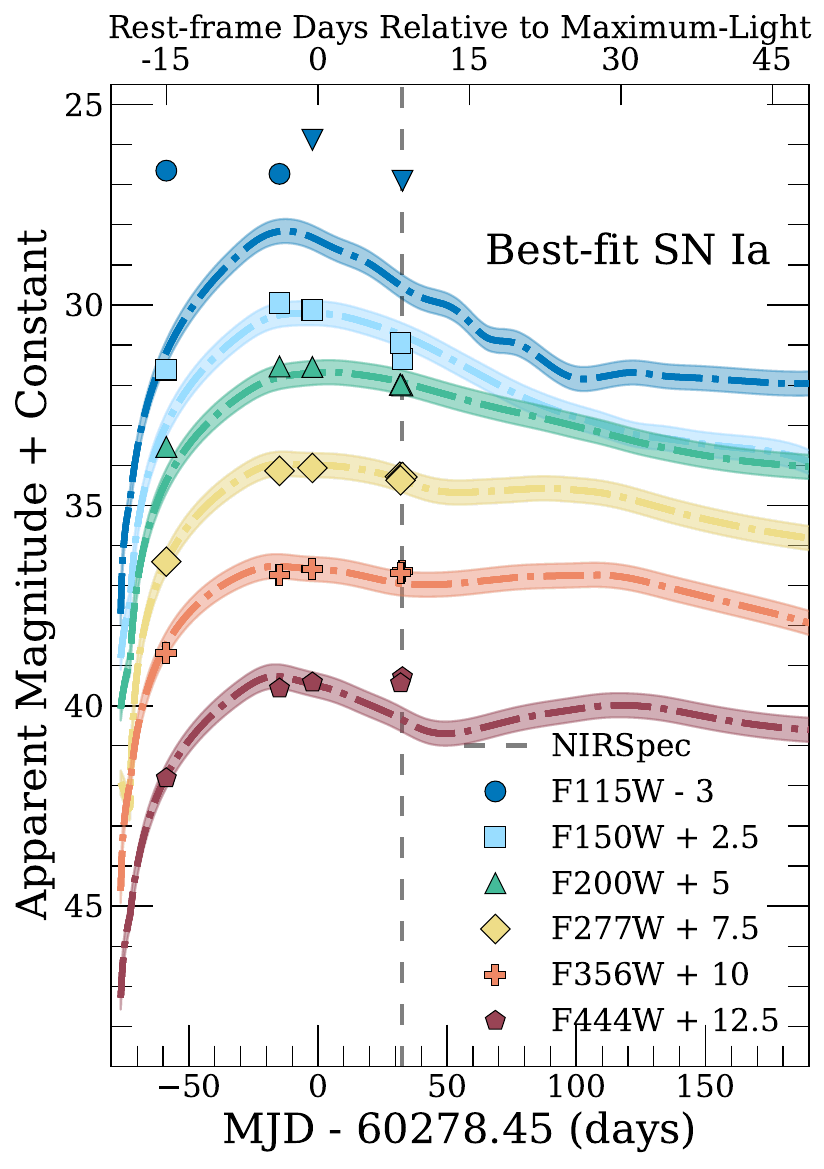}
    \caption{(\textit{left}): The multiband light curve of \highzIc (colored points), plotted with the offsets shown for visual clarity, and its best-matched SN~Ib/c light curve (dashed-curves). The epoch of our NIRSpec Prism spectrum is marked with the dashed-vertical line, and indicates an approximate phase of +11 days relative to peak brightness in \textit{F150W}. (\textit{middle}): Similar to the left panel but showing the best-fit SN~II. (\textit{right}): Similar to the other panels but showing the best-fit SN~Ia SALT3-NIR model \citep{Pierel22}, with shape and color parameters of $x_1 = -0.29$ and $c = 0.63$, respectively.\label{fig:phot}
    }
\end{figure*}

We examine the light curve of \highzIc (originally presented in \citetalias{Decoursey24}) in \autoref{fig:phot}. The data shown (colored points) are NIRCam wideband photometry. We fit the observed photometry using all existing core-collapse, SN light curve evolution models with rest-frame optical to near-IR (to observer-frame $\sim4\mu$m) wavelength coverage \citep{pierel_extending_2018}, and the SALT3-NIR SN\,Ia light curve model \citep{Pierel22}. Prior to fitting, we correct for Galactic extinction ($E(B-V)=0.01$ mag with $R_V=3.1$). In \autoref{fig:phot}, we show the best-fit SN~Ib/c model, SN~II model, and SALT3-NIR SN Ia model from left to right. The epoch of our NIRSpec DDT observation is represented by the vertical-dashed line. For the SN~Ib/c model, this occurs at a phase of $+11$ days relative to peak-brightness in F150W ($+14$d and $+8$d for the SN~II and SN~Ia model, respectively). For the SN~Ia model, we find best-fit shape and color parameters of $x_1 = -0.29$ and $c = 0.63$, respectively. The core-collapse light curve models do not have uncertainties. Therefore, to determine relative goodness of fit, we add a constant uncertainty to each model light curve such that $\chi^{2}/\nu = 1.0$ for the best-fit (SN~Ib/c). We add identical uncertainty to the SN II and SALT3-NIR SN~Ia models so that the $\chi^{2}/\nu$ values are directly comparable. This uncertainty is shown by the shaded regions in \autoref{fig:phot}.  The model parameters for each fit are shown in \autoref{tab:lcfit}. 

We find that SESNe reproduce the multiband light curve evolution of \highzIc. Specifically, from each of the top 10 core-collapse SN~Ib/c and SN~II fits, we determine $\chi^{2}/\nu = 1.6\pm0.4$ and $ 2.4\pm0.5$, for each SN type, respectively. The best-fit SN~Ia model has $\chi^{2}/\nu = 1.75$. While all fits perform well in the rest-frame optical (F150W, F200W, F277W, and F356W), the best-fit SN~Ia is unable to reproduce the flux at rest-frame NUV wavelengths (F115W). The best-fit SN~Ib/c and SN~II are also better matched to the last epoch in F444W. Due to limitations of the data, we caution that the light curves models are most uncertain at the reddest wavelengths. If we assume $H_{0} = 70$~km~s$^{-1}$~Mpc$^{-1}$ \citep{Riess16, Riess18:gaia}, we find a high peak luminosity of $M_V$ = $-19.0 \pm 0.2$ mag ($M_R$ = $-19.1 \pm 0.2$ mag) for the SN~Ib/c model. We note that the best-fit SN II models similarly require high-peak brightnesses ($M_V$ = $-19.4 \pm 0.3$ mag) that are difficult to reconcile with the observed luminosity functions of normal-SN~II \citep{Li11:rate2,valenti16}. Similar to $\chi^{2}/\nu$, we derive these uncertainties from the distribution of the 10 best-fit SN~Ib/c models. 

\begin{figure}[htb!]
    \centering
    \includegraphics[width=3.2in]{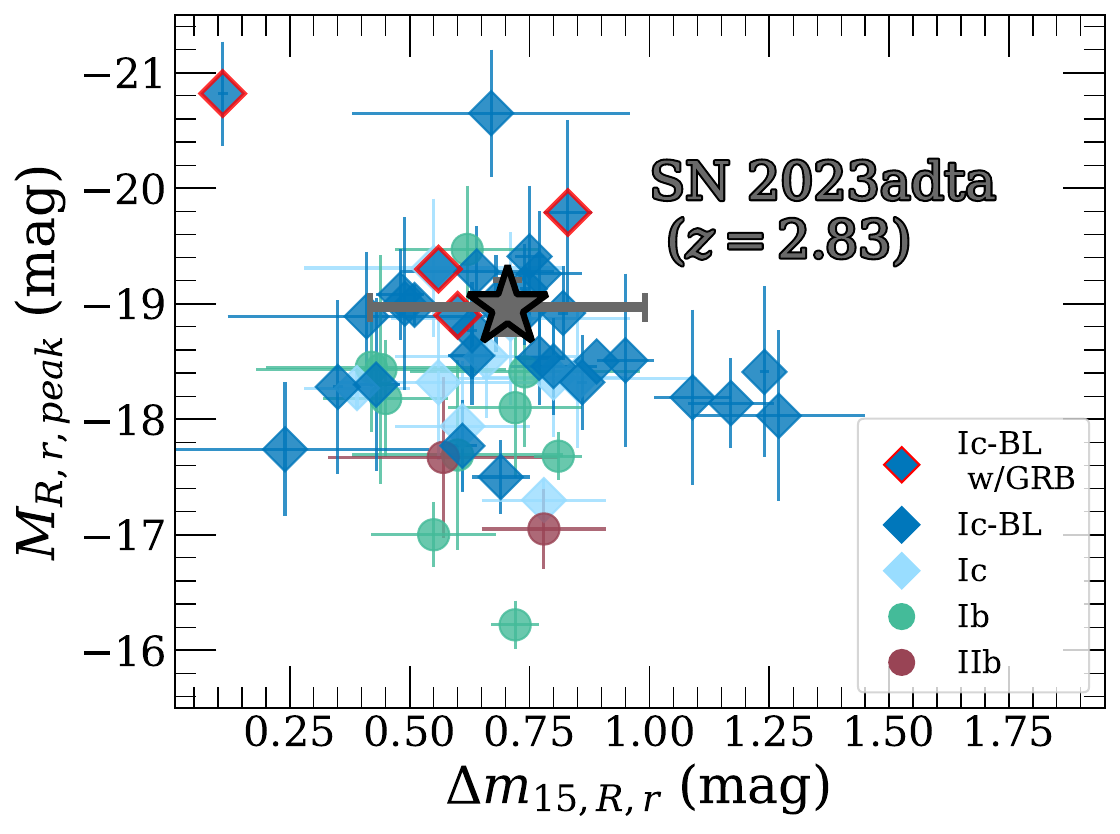}
    \caption{Comparison of the light curve peak brightness ($M_{R,r}$) and shape ($\Delta m_{15,R,r}$) characteristics of \highzIc to other SESNe from \citet{Drout11} and SNe~Ic-BL from \citet{Taddia19}. Measurements were derived from $R$- and $r$-band photometry, respectively.
    \label{fig:dm15}}
\end{figure}

Some studies have shown that SN Ic-BL light curves are very similar to those of other SNe Ib/c but tend toward higher absolute luminosities \citep{Drout11, Taddia15, Prentice16, Lyman16}. In particular, \citet{Drout11} found a mean $M_{R,peak} = -19.0 \pm 1.1$ mag for SNe~Ic~BL ($M_{R,peak} = -17.9 \pm 0.9$ mag and $-18.3 \pm 0.6$ mag for SNe~Ib and SNe~Ic, respectively). In \autoref{fig:dm15}, we show a comparison of the \highzIc peak brightness ($M_R$) and light curve $\Delta m_{15,R}$ shape, to other SESNe from \citet{Drout11} and SNe~Ic-BL from \citet{Taddia19}. These measurements were derived from $R$- and $r$-band photometry, respectively. Given its sparsely-sampled light-curve, $\Delta m_{15,R}$ is uncertain but consistent with other SESNe. Its peak brightness is similar to those of other SN~Ic-BL, in particular, those with associated GRBs (blue diamonds with red edges). Using the relationship derived in \citet{Drout11}:
\begin{equation}
    log(M_{Ni}) \approx - 0.41 M_R - 8.3 \\
\end{equation}
we estimate that $M_{Ni} = 0.3 \pm 0.1$ $M_{\odot}$ for \highzIc.

\subsection{Spectroscopic Classification \& Comparisons}\label{s:spec_comp}

\begin{figure*}[htb!]
    \centering
    \includegraphics[width=6.2in]{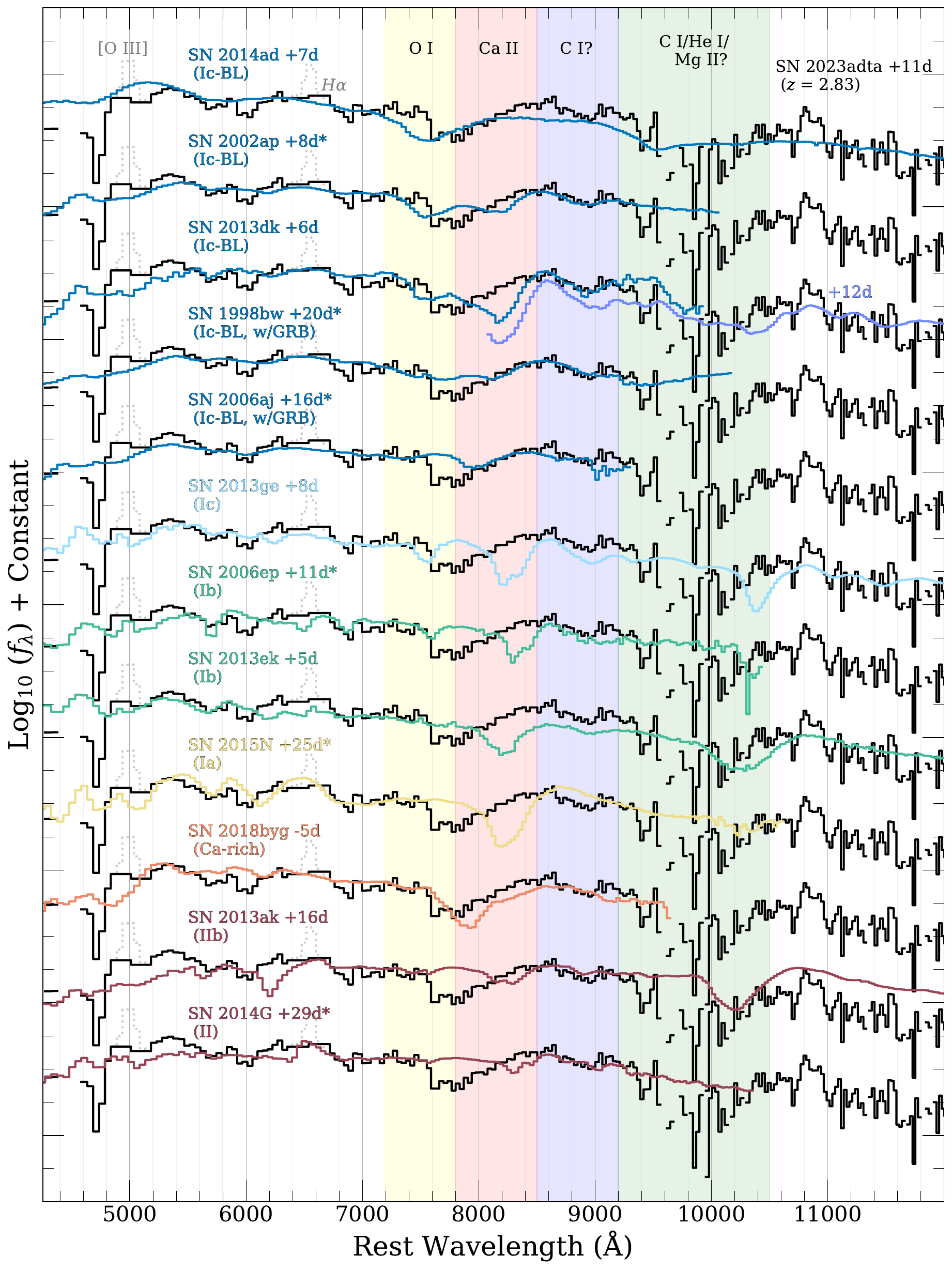}
    \caption{Rest-frame NIRSpec Prism spectrum of \highzIc (black curves) compared with spectra of a variety of other SN types at a similar phases (colored curves). We mask negative values that result from imperfect host-galaxy subtraction. The dotted-gray curves show clipped host galaxy emission lines. The color of each other curve corresponds to the SN type. From top to bottom we show: five SNe~Ic-BL (blue), one SN~Ic (light-blue), two SNe~Ib (green), one SN~Ia (yellow), one Ca-rich SN (orange), and one SN~IIb (dark red). A ``*" next to the phase of a comparison spectrum indicates that it was a matched spectrum using the SN classification code {\tt NGSF} \citep{Goldwasser22}.
    \label{fig:class}}
\end{figure*}

\begin{deluxetable}{clcccc}
    \tablecaption{Best-fit {\tt{NGSF}} parameters for \highzIc \label{tab:specfit}. Below the double-line we show the highest rankings for specific SN types.\label{tab:ngsf}}
    \tablehead{\colhead{Rank} & \colhead{SN} & \colhead{Type} & \colhead{Phase (days)} & \colhead{$A_V$ (mag)} & \colhead{$\chi^2/\nu$}}
    \tablecolumns{6}
        \startdata
        1&SN~2006aj&Ic-BL&+16&0.45&3.19\\
        2&SN~2002ap&Ic-BL&+8&1.25&3.42\\
        3&SN~1998bw&Ic-BL&+20&1.90&3.52\\
        4&PTF10vgv&Ic-BL&+2&2.0&3.93\\
        5&SN~2014G&II&+29&1.2&4.12\\
        \hline
        \hline
        8&SN~1994L&Ic&+2&2.0&4.37\\
        10&SN~2013df&IIb&+14&1.0&4.45\\
        11&SN~2015N&Ia&+25&0.0&4.57\\
        21&SN~2006ep&Ib&+11&0.0&5.18\\
        \enddata
\end{deluxetable}

\begin{figure}[htb!]
    \centering
    \includegraphics[width=2.2in]{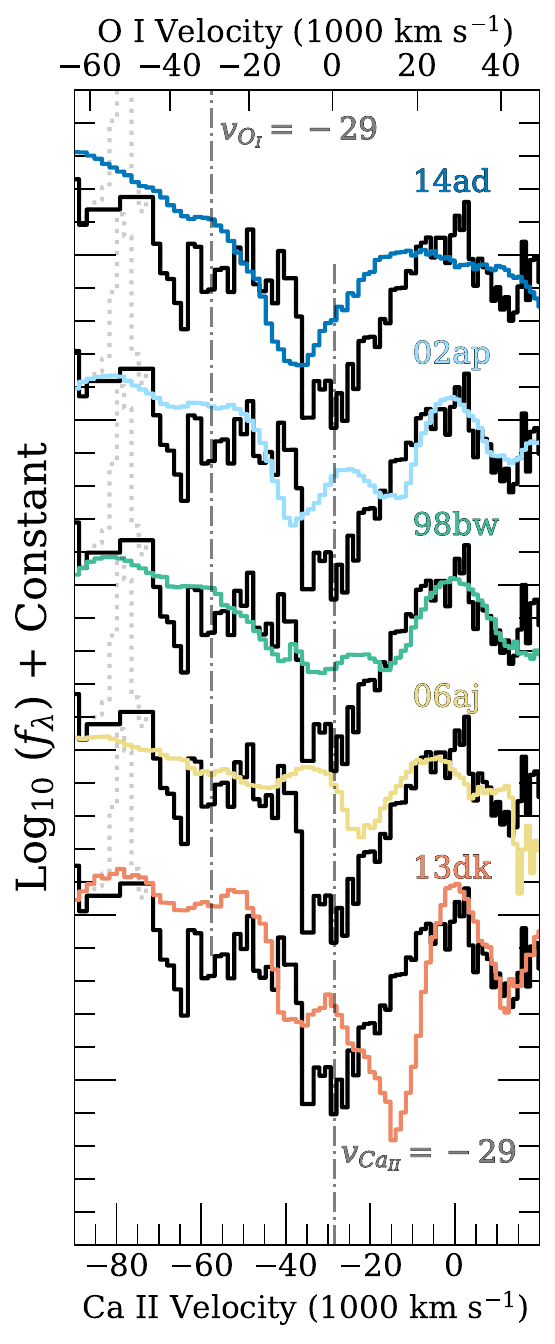}
    \caption{Similar to \autoref{fig:class} but zoomed in on the \ion{O}{1} and \ion{Ca}{2} NIR triplet region of each SN~Ic-BL. The velocity width of the primary absorption feature suggests ejecta moving at $>20{,}000$ \kms. The velocity we measure from the absorption minimum is  $29{,}000 \pm 2{,}000$ \kms.
    \label{fig:caoi}}
\end{figure}

The low resolution of the NIRSpec Prism and uncertainty in underlying contamination from host galaxy light provide additional challenges in the classification of \highzIc. The presence of strong-host galaxy $H{\alpha}$ emission with FWHM  = $4{,}000$ \kms could bias classification against SNe~II, especially those with relatively narrow emission features like SNe~IIn. 

The spectroscopic classification of \highzIc was performed with Next Generation SuperFit\footnote{\url{https://github.com/oyaron/NGSF}} \citep{Goldwasser22} based on the IDL code Superfit \citep{Howell05}. This code allows for the simultaneous fitting of the host galaxy  light contamination and SN flux. We restrict the fit to use only post-maximum-light template SNe and the ``SB1" host galaxy template which best-matches the continuum of our host galaxy spectrum (see \autoref{fig:2d1d}). We also restrict the extinction parameter to $0 < A_V < 2$ mag. We de-redshift the SN spectrum, clip prominent host-galaxy emission lines, and then perform {\tt NGSF} fitting at a resolution of $30$ \AA. This is similar to the native wavelength binning output by the \textit{JWST} Stage 3 pipeline which ranges from $25 - 52$ \AA\ across our de-redshifted spectrum of \highzIc. 

A summary of the spectroscopic fitting results is presented in \autoref{tab:ngsf}. Ordered by $\chi^2/\nu$, the four best-fit SN spectra are from SNe~Ic-BL with phases ranging from $+2$ to $+20$ days, in good agreement with the results of our light curve fit ($+11$d). Interestingly, two of these SNe (SN~1998bw and SN~2006aj) have associated GRBs. In each of these fits, the SN contributions are $>98\%$ indicating little host contamination. For SNe~Ic-BL, {\tt{NGSF}} prefers a high amount of reddening ($A_V > 0.45$ mag), which is consistent with our derived host galaxy parameters ($A_V = 1.8$ mag, see \autoref{s:host}), but not our best-fit light curve ($A_V = 0$ mag). However, we caution that imperfect subtraction of the host galaxy light could strongly influence this parameter. Similar to the light curve fitting, the second best fit SN type is SN~II, however, its phase ($+29$d) is significantly greater than what we would expect based on best-fit SN~II light curve ($+13$d). 

Given the limited SN template bank available in {\tt NGSF}, additional comparisons are needed to provide a secure classification for \highzIc. We show spectroscopic matches to \highzIc from {\tt NGSF} along with other useful comparison spectra in \autoref{fig:class}. For log$(f_{\lambda})$ comparisons, we mask negative values in \highzIc that result from imperfect host-galaxy subtraction.

From top to bottom, we show our NIRspec Prism spectrum of \highzIc in comparison to five SNe~Ic-BL (SN~2014ad, \citealt{Kwok22, Shahbandeh22}; SN~2002ap, \citealt{Mazzali02}; SN~2013dk, \citealt{EliasRosa13, Shahbandeh22}; SN~1998bw, \citealt{Galama+98Natur}; and SN~2006aj, \citealt{Pian06}), one SN~Ic (SN~2013ge, \citealt{Drout16, Shahbandeh22}), two SNe~Ib (SN~2006ep, \citealt{Smartt15_pessto}; and SN~2013ek, \citealt{Smartt15_pessto, Shahbandeh22}), one SN~Ia (SN~2015N, \citealt{Stahl20}), one Ca-rich SN (SN~2018byg, \citealt{De19}), and one SN~IIb (SN~2013ak, \citealt{Smartt15_pessto, Shahbandeh22}). Four of these SNe (SN~2014ad, SN~2013ge, SN~2013ek, and SN~2013ak) have optical spectra that were observed within one day of a NIR spectrum presented in \citet{Shahbandeh22}. In these cases, we have combined these data into one full optical-to-NIR spectrum. All comparison spectra have been re-binned to a have a similar wavelength spacing to the NIRSpec spectrum of \highzIc. We mark spectroscopic fits from {\tt NGSF} with a ``*''.

There are two primary broad absorption features in the spectrum of \highzIc that agree relatively well with other SNe~Ic-BL, one at $\sim 8{,}000$ \AA\ and another at $\sim 10{,}000$ \AA. The former is likely produced by the \ion{Ca}{2} NIR triplet, and the latter is likely produced from blended contributions from \ion{C}{1}, \ion{He}{1}, and \ion{Mg}{2} \citep{Shahbandeh22}. Absorption from \ion{O}{1} and \ion{Ca}{2} can appear as separate features in normal SNe~Ic and SNe~Ib due to their lower velocities \citep{Matheson+01}. However, in SNe~Ic-BL, these features can blend together due to the larger ejecta velocities \citep{Foley03}. SN~Ic-BL comparison spectra in \autoref{fig:class} have been selected to showcase the diversity in this wavelength region. Additionally, the spectrum of SN~2013dk at +6 days (+12 in the NIR) exhibits some interesting similarity to \highzIc at $\sim 9{,}000$ \AA. \citet{Shahbandeh22} identified this absorption feature as \ion{C}{1} in SN~2013dk. This SN also matches the fall-off in  flux near $4{,}000$ \AA\ better than most of the other comparison SNe. 

We also make spectroscopic comparisons to normal SESNe and a SN~Ia. Generally, these SNe are unable to reproduce the large blueshifts seen in \highzIc and the line strengths relative the continuum of these SNe are not well matched. The presence of \ion{He}{1} and/or \ion{Si}{2} in the spectrum of \highzIc is ambiguous. \highzIc has additional shallow absorption features near $6{,}000$ \AA\ that could result from these elements but the resolution of the data, and uncertainty on phase make a definitive line identification in this wavelength region very difficult. Furthermore, \ion{He}{1} $\lambda 10830$~\AA\ could contribute to the broad absorption at $10{,}000$~\AA, however, there is no way to differentiate from potential contributions from \ion{C}{1} $\lambda 10693$~\AA\, and/or \ion{Mg}{2} $\lambda 10927$~\AA\ \citep{Shahbandeh22}. Furthermore,  For the SN~Ia (yellow curve, SN~2015N), {\tt NGSF} prefers a later phase of $+25$ days. This could be due to the color of \highzIc at the epoch of our spectrum. A SN~Ia at an earlier phase as suggested by the light curve fit, would likely be too blue to reproduce the observed continuum. 

Some peculiar transients like Ca-rich SNe are red at early times due strong line-blanketing \citep{Shen18a,Polin19, Jacobson-Galan20}. In \autoref{fig:class}, we show one such event, SN~2018byg (orange curve, \citealt{De19}), at $-5$ days relative to maximum-light. This spectrum reproduces the continuum of \highzIc well and also has a high \ion{Ca}{2} velocity of $-22{,}000$ km s$^{-1}$ \citep{De19,Jacobson-Galan20}, however, the strength of this absorption does not match as well as the \ion{Ca}{2} features in our SNe~Ic-BL comparison spectra. Finally, we compare to a SN~IIb and a SN~II (dark red curves, SN~2013ak and SN~2014G). These SNe have low ejecta velocities that are difficult to reconcile with \highzIc. We measure a H$\alpha$ FWHM$=4{,}000$ \kms which is consistent with the resolution of the NIRSpec Prism at this wavelength indicating that this emission is from the host galaxy. However, a clear continuum is difficult to define in this region and broader H$\alpha$ may be present.

Even though there are significant caveats related to both the photometric and spectroscopic classification methods, the observational characteristics of \highzIc outlined below provide convincing evidence that it can be classified as a SN~Ic-BL at z=\snz: 

\begin{itemize}
  \item Light curve fit strongly prefers SN~Ib/c which reproduces the \textit{F115W} flux at early epochs. 
  \item {\tt NGSF} fitting of the NIRSpec spectrum strongly suggests the subclass of SN~Ic-BL at a phase consistent with the light curve. 
  \item Broad, high-velocity absorption features are inconsistent with those observed in normal SNe Ia, Ib, and Ic.
  \item Lack of obvious H or He absorption/emission features.
  \item Red continuum shape and flux drop-off at rest-frame $4000$ \AA. 
\end{itemize}

\subsection{Ejecta Velocity and Physical Parameters}\label{s:phys}

\begin{deluxetable}{cccc}
    \tablecaption{Estimated explosion parameters for \highzIc\label{tab:phys}.}
    \tablehead{\colhead{$v_{ph}$ (\kms)} & \colhead{$M_{Ni}$ $(M_{\odot})$} & \colhead{$M_{ej}$ $(M_{\odot})$} & \colhead{$E_{KE}$ ($10^{51}$ erg)}}
    \tablecolumns{6}
        \startdata
        $29{,}000 \pm 2{,}000$&$0.3 \pm 0.1$&$5^{+4}_{-2}$&$24^{+29}_{-14}$\\
        $20{,}000$&$0.3 \pm 0.1$&$3^{+3}_{-2}$&$8^{+11}_{-5}$\\
        $10{,}000$&$0.3 \pm 0.1$&$1.6^{+1.8}_{-0.9}$&$1.0^{+2.0}_{-0.7}$
        \enddata
\end{deluxetable}

Using a combination of both photometric and spectroscopic properties of \highzIc, we can derive some physical parameters of the explosion. \citet{Drout11} showed how peak luminosity ($M_R$), light curve shape ($\Delta m_{15,R}$), and photospheric velocity ($v_{ph}$) are related to the nickel mass ($M_{Ni}$), ejecta mass ($M_{ej}$), and kinetic energy of the explosion ($E_{KE}$). To get an estimate of $v_{ph}$ we examine the \ion{Ca}{2} and \ion{O}{1} absorption features of SNe~Ic-BL in more detail in \autoref{fig:caoi}. The spectrum of SN~2014ad reproduces the shape of the primary absorption features well but at larger blueshifts. \citet{Kwok22} attribute the absorption at $\sim 7{,}000$ \AA\ primarily to \ion{O}{1} and estimate a photospheric velocity of $22{,}000$ km s$^{-1}$ via a best-fit TARDIS \citep{Kerzendorf14} model. 

Assuming the absorption minimum in the spectrum of \highzIc is primarily caused by \ion{Ca}{2}, using an intensity-weighted rest wavelength of 8567~\AA\ for the \ion{Ca}{2} NIR triplet, we estimate an ejecta velocity of $-29{,}000 \pm 2{,}000$ km s$^{-1}$ which is consistent with other SN~Ic-BL \citep{Modjaz+16,Taddia19}. We find that \highzIc may have another weaker absorption feature blueward of \ion{Ca}{2}, which if caused by \ion{O}{1}, also indicates ejecta with velocities ranging from $20{,}000$ to $40{,}000$ \kms. An estimate of the \ion{Fe}{2} $5169$ \AA\ line velocity would be a better indicator of the photospheric velocity since this line does not saturate \citep{Branch02, Modjaz+16}, however, the strong residual [\ion{O}{3}] emission from the host galaxy prohibits our ability to make this measurement. 

Using the properties of the light curve that we estimated in \autoref{s:lc}, we derive physical explosion parameters for \highzIc using the equations 1 and 2 and methods described in \citet{Drout11}. We determine these parameters using our measured \ion{Ca}{2} velocity as the photospheric velocity, but also calculate results for fiducial velocities of $20{,}000$ \kms and $10{,}000$ \kms for comparison. These parameters are shown in \autoref{tab:phys}. We estimate a large ejecta mass and kinetic energy of $5^{+4}_{-2}$ $M_{\odot}$ and $24^{+29}_{-14}\times 10^{51}$ ergs, respectively. The large uncertainties primarily arise from the poor constraints on the light curve shape and strong dependence ejecta velocity. Nevertheless, these estimates are consistent with what we expect for engine driven explosions \citep{Drout11, Cano13,Taddia19}.

\subsection{Host Galaxy Properties}\label{s:host}


\begin{figure*}[htb!]
    \centering
    \includegraphics[width=5.5in]{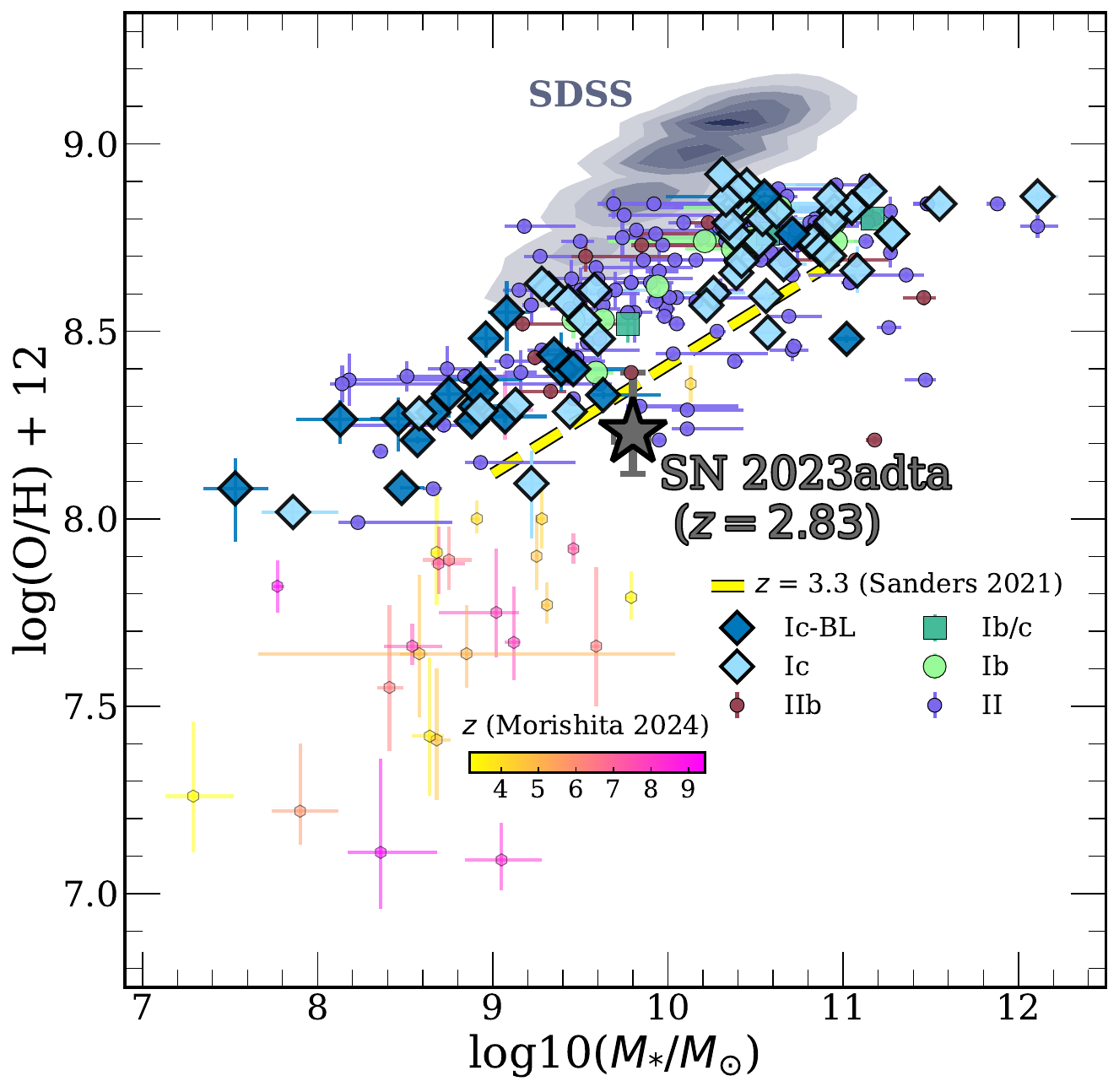}
    \caption{A comparison of the host galaxy properties of \highzIc to the mass-metallity relationship observed in SDSS field galaxies (gray contours, $z\sim0$, \citealt{Eisenstein11}), core-collapse SN host galaxies from \citet{Kelly12} and additional SN~Ic/Ic-BL host galaxies from \citet{Modjaz20} (colored points), and high-$z$ galaxies observed with \textit{JWST} (pink-yellow hexagons, \citealt{Morishita24}). The dashed-yellow line is the empirical, $z=3.3$, mass-metallicity relationship derived in \citet{Sanders21}. 
    \label{fig:MZ}}
\end{figure*}

The presence of a SN~Ic-BL at $z=\snz$ provides a unique opportunity to study the environment of this rare subclass of core-collapse supernova in the early Universe. Its host galaxy is irregular with several bright knots of star-formation (see \autoref{fig:diff}). We derive galaxy properties by performing a {\tt BAGPIPES} \citep{Carnall18,Carnall19} fit to the host galaxy spectrum that was flux-calibrated to the total host flux. From this fit we find mass log$_{10}(M_{*}/M_{\odot}) = 9.78^{+0.09}_{-0.11}$,  stellar metallicity $Z = 0.35^{+0.16}_{-0.08} Z_{\odot}$, $A_V = 1.832^{+0.10}_{-0.14}$, timescale $\tau = 2.10^{+0.06}_{-0.05}$ Gyr (SFR$\propto e^{t/\tau}$), and redshift $z = 2.833^{+0.002}_{-0.003}$. Assuming a solar metallicity of log(O/H) + 12 = 8.69 \citep{Asplund09}, we convert the best-fit $Z$ from {\tt{BAGPIPES}} to oxygen abundance and find log(O/H) + 12 =  $8.23^{+0.16}_{-0.12}$. We caution that this conversion assumes that stellar and gas-phase metallicity are equal. 

We note that the redshift uncertainty from this fit is a factor of two larger than from the Gaussian fitting method discussed in \autoref{s:red}. This method used the highest S/N emission lines that were close in wavelength relative to the wavelength range of the NIRSpec PRISM. Since {\tt{BAGPIPES}} fits over the full wavelength range, there may be additional uncertainties in the wavelength solution that are getting factored into the redshift uncertainty. Furthermore, the NIRSpec PRISM resolution varies by a factor of $\sim$3 over the entire wavelength range. Since this widens the line spread function at lower wavelengths, the inclusion of these regions in the fit could also lead to the larger observed uncertainty.

In \autoref{fig:MZ}, we compare the mass and metallicity of the host galaxy of \highzIc to the mass-metallicity relation of low-$z$ galaxies (SDSS DR8, \citealt{Eisenstein11}), high-$z$ galaxies ($z\sim 3 - 9$, \citealt{Curti23, Nakajima23, Morishita24}), core-collapse supernova host galaxies \citep{Kelly12, Modjaz20}. These data are presented as the gray contours, pink-yellow hexagons, and colored points, respectively. The SDSS and SN host galaxy metallicities were derived via the PP04 O3N2 calibration \citep{Pettini04}, and metallicities of the high-$z$ galaxies were derived using the direct electron-temperature ($T_e$) method. We also compare to the mass-metallicity relationship derived for galaxies at $z\sim 3.3$ in the MOSDEF survey (dashed-yellow line, \citealt{Sanders21}). 


We find that the metallicity of the host galaxy of \highzIc is in good agreement with those of other SN~Ic-BL host galaxies (dark blue diamonds). This corroborates previous work that has determined that these high-energy events tend to occur in lower-metallicity environments than those of other SESNe and normal core-collapse SNe \citep{Sanders12, Kelly12, Modjaz20}. In particular, our measured metallicity for \highzIc, is very similar to the mean SN~Ic-BL PP04 O3N2 metallicity, log(O/H) + 12 =  $8.31\pm0.04$, and mean SN-GRB PP04 O3N2 metallicity log(O/H) + 12 =  $8.20\pm0.06$, reported in \citet{Modjaz20}. Furthermore, both the mass and metallicity of the host galaxy of \highzIc are consistent with what we expect for galaxies at this redshift. Given their rarity ($\sim 4$\% the rate of core-collapse SNe by volume, \citealt{Shivvers17:loss}), the likelihood of observing a SN~Ic-BL in the JADES DDT sample is small. However, since the metallicities of high-$z$ galaxies are lower, it is possible that SNe~Ic-BL occur at an enhanced rate at higher redshifts. We note that the mean galaxy mass of the SN~Ic-BL host galaxies displayed is log$_{10}(M_{*}/M_{\odot}) = 9.1$ with a standard deviation of $0.8$. \highzIc has relatively high a host galaxy log$_{10}(M_{*}/M_{\odot}) = 9.78^{+0.09}_{-0.11}$. However, \highzIc was not discovered in an untargeted survey like the majority of SNe~Ic-BL displayed. Since the presence of a clear host galaxy was used to motivate its spectroscopic follow-up, there could be a significant selection effect in the JADES SN sample for SNe in more massive host galaxies.

\subsection{Searching for Coincident GRBs with \highzIc}\label{s:grb}

The connection between SNe~Ic-BL and LGRBs would be interesting to constrain at high-$z$. We used the online database GRBWeb\footnote{\url{https://user-web.icecube.wisc.edu/~grbweb_public/index.html}} to search for spatially and temporally coincident GRBs that could be associated with \highzIc. We find that the best candidate for an associated GRB is GRB230911A. The MJD of this event was 60198, 66 days prior to our first detection of \highzIc. Based on our best-fit light curve, this GRB occurred at a phase of $-18$d. The location of this GRB (RA = 59.8, Dec = -34.4, GCN Circular 34652\footnote{\url{https://gcn.nasa.gov/circulars/34652}}) is 4.3 degrees from \highzIc with an uncertainty of 4.1 degrees. However, GRB230911A also has an associated optical afterglow GOTO23akf/AT2023shv \citep{Belkin24} at $=3$h$50$m$0.51$s decl. $=-29$d$49$m$30.66$s with 0.3 arcsecond uncertainty. With a separation of 4.3 degrees from \highzIc, we conclude GRB230911A is not associated with this SN. Furthermore, a \textit{Swift}-XRT afterglow was also detected (GCN Circular 34702 \footnote{\url{https://gcn.nasa.gov/circulars/34702}}) at R.A. $=3$h$50$m$0.65$s decl. $=-29$d$49$m$33.2$s with 3.4 arcsecond uncertainty, providing additional support for conclusion. We do not find any other GRBs that are likely to be spatially and temporally coincident with \highzIc.

\section{Discussion \& Conclusions}\label{s:conc}

As part of a DDT follow-up program for the JADES Transient Survey (\citetalias{Decoursey24}), we used \textit{JWST} NIRCam and NIRSpec observations to confirm the presence of a SN~Ic at $z=\snz$ (\highzIc). To date, this is the most distant SESN ever discovered, occurring when the age of the universe was approximately 2.2 Gyr. Analysis of the light curve and spectrum of \highzIc, provide strong evidence that this SN fits within the SN~Ic-BL subclass of SESNe. These SNe are characterized by their broad absorption features indicating typical ejecta velocities of $\sim20{,}000 - 30{,}000$\kms, high explosion energies, lack of H and He features, and have been associated with LGRBs. This subclass also has a preference for low-metallicity environments relative to other SESNe \citep{Sanders12, Kelly12, Modjaz20}. 

\highzIc satisfies many of these criteria. First, we fit the observed light curve and found a best-fit photometric classification of SN~Ib/c. In particular, this model reproduces the flux in F115W better than any other SN type. Our best fit indicates that the phase of NIRSpec observation is at $\sim +11$ days. While it difficult to distinguish SN~Ib/c/c-BL using solely the light curve shape, we find that \highzIc had a peak luminosity in $M_V$ = $-19.0 \pm 0.2$. This is consistent with the high-luminosities observed in SNe~Ic-BL relative to other SESNe \citep{Drout11, Cano13, Taddia19}, and the high peak luminosity of the best-fit SN~II models, $M_V$ = $-19.4 \pm 0.3$, is inconsistent with the luminosity distribution of normal-SN~II. Furthermore, engine-driven explosions, the subset of SN~Ic-BL characterized by their associations with LGRBs, tend to have the highest luminosities within the SN~Ic-BL subclass \citep{Drout11,Taddia19}. 

Second, we performed spectroscopic classification of \highzIc on our NIRSpec Prism spectrum using the fitting code {\tt NGSF} based on the IDL code Superfit \citep{Howell05}. This code models contributions from the SN and host galaxy and found results consistent with our photometric classification. \highzIc has broad absorption features in the optical and NIR that are consistent with those observed in SNe~Ic-BL. The absorption feature at a rest wavelength of $\sim 8{,}000$ \AA\ is likely produced by the \ion{Ca}{2} NIR triplet, and implies ejecta velocities of $\sim 29{,}000$ \kms, in agreement with the large explosion energy that we infer from the light curve. Additionally, we do not find strong evidence for H or He features that would be typical for normal SNe~Ib or SNe~II. In combination with the light curve properties of \highzIc, we estimate nickel mass, ejecta mass and kinetic energy ($0.3\pm0.1$ $M_{\odot}$, $5^{+4}_{-2}$ $M_{\odot}$ and $24^{+29}_{-14}\times 10^{51}$), that are consistent with typical values for SNe~Ic-BL \citep{Taddia19}. 

Third, we estimated properties of the host galaxy of \highzIc and found that its metallicity is low relative to other core-collapse SNe observed in the local Universe. This is in agreement with previous analyses of SNe~Ic-BL host galaxies and what is expected for the galaxy mass-metallicity relationship at $z=\snz$. Lastly, we do not identify any GRBs that are coincident with \highzIc.

SN~Ic-BL are rare in the local universe ($\sim4$\% of the core-collapse SN rate, \citealt{Shivvers17:loss}), therefore, detection of a SN~Ic-BL at $z=\snz$ within a small sample suggests that the rate of these events may be enhanced at this distance. Future observations are needed to constrain the rates of these SNe at high-$z$ and to understand how their host populations may or may not evolve with redshift.

\textit{JWST} is a remarkable tool for this task and in general, the characterization of the high-$z$ transient Universe. The NIRSpec MOS mode is capable of observing several candidate SNe at once, making it very efficient for transient follow-up and classification. The work presented here highlights some additional challenges that the community will face as a sample of high-$z$ SNe continues to be constructed. Specifically, at these distances it can become very difficult to separate light from a SN and its host galaxy. Host galaxy continuum contamination can make SN features appear weak and difficult to distinguish, and at low-resolution (like the NIRSpec Prism), galaxy emission line contamination can preclude the identification of key SN features (e.g., broad H$\alpha$). Future high-$z$ transient survey design should consider these limitations and take steps to mitigate their impact. Furthermore, the \textit{Nancy Grace Roman Space Telescope} will also include only a low-resolution dispersive element and SN classification will be similarly difficult. More work is needed to improve classification at low-resolution in order to get the most out of future high-$z$ SNe observed with these telescopes. 

\begin{center}
    \textbf{Acknowledgements}
\end{center}

This work is based on observations made with the NASA/ESA/CSA \textit{JWST} as part of programs 1180 and 6541. This paper is based on observations with the NASA/ESA Hubble Space Telescope and James Webb Space Telescope obtained from the Mikulski Archive for Space Telescopes at STScI. We thank the DDT and JWST/HST scheduling teams and instrument experts at STScI for extraordinary effort in getting the DDT observations used here scheduled quickly. 

MRS is supported by an STScI Postdoctoral Fellowship. JDRP is supported by NASA through a Einstein Fellowship grant No. HF2-51541.001 awarded by the Space Telescope Science Institute (STScI), which is operated by the Association of Universities for Research in Astronomy, Inc., for NASA, under contract NAS5-26555. This research is based (in part) on observations made with the NASA/ESA Hubble Space Telescope obtained from the Space Telescope Science Institute, which is operated by the Association of Universities for Research in Astronomy, Inc., under NASA contract NAS 5–26555. 
Part of the JWST data used in this paper can be found in MAST: \dataset[10.17909/8tdj-8n28]{https://dx.doi.org/10.17909/8tdj-8n28} (JADES DR1). 

Additionally, this work made use of the {\it lux} supercomputer at UC Santa Cruz which is funded by NSF MRI grant AST 1828315, as well as the High Performance Computing (HPC) resources at the University of Arizona which is funded by the Office of Research Discovery and Innovation (ORDI), Chief Information Officer (CIO), and University Information Technology Services (UITS).
AJB acknowledges funding from the “FirstGalaxies” Advanced Grant from the European Research Council (ERC) under the European Union’s Horizon 2020 research and innovation program (Grant agreement No. 789056). 

PAC, EE, DJE, BDJ, are supported by JWST/NIRCam contract to the University of Arizona, NAS5-02015.
DJE is also supported as a Simons Investigator.
RM acknowledges support by the Science and Technology Facilities Council (STFC), by the ERC through Advanced Grant 695671 “QUENCH”, and by the UKRI Frontier Research grant RISEandFALL. RM also acknowledges funding from a research professorship from the Royal Society.

BER acknowledges support from the NIRCam Science Team contract to the University of Arizona, NAS5-02015, and JWST Program 3215. JDRP is supported by NASA through a Einstein Fellowship grant No. HF2-51541.001 awarded by the Space
Telescope Science Institute (STScI), which is operated by the
Association of Universities for Research in Astronomy, Inc.,
for NASA, under contract NAS5-26555. 
YZ thanks Matteo Cantiello for helpful discussions.

\facilities{JWST (NIRCam/NIRSpec)}

\software{astropy \citep{astropy,astropy2,astropy3}}

\newpage
\bibliography{astro_refs}

\begin{thebibliography}{}
\expandafter\ifx\csname natexlab\endcsname\relax\def\natexlab#1{#1}\fi
\providecommand{\url}[1]{\href{#1}{#1}}
\providecommand{\dodoi}[1]{doi:~\href{http://doi.org/#1}{\nolinkurl{#1}}}
\providecommand{\doeprint}[1]{\href{http://ascl.net/#1}{\nolinkurl{http://ascl.net/#1}}}
\providecommand{\doarXiv}[1]{\href{https://arxiv.org/abs/#1}{\nolinkurl{https://arxiv.org/abs/#1}}}

\bibitem[{{Anderson} {et~al.}(2018){Anderson}, {Dessart}, {Guti{\'e}rrez}, {Kr{\"u}hler}, {Galbany}, {Jerkstrand}, {Smartt}, {Contreras}, {Morrell}, {Phillips}, {Stritzinger}, {Hsiao}, {Gonz{\'a}lez-Gait{\'a}n}, {Agliozzo}, {Castell{\'o}n}, {Chambers}, {Chen}, {Flewelling}, {Gonzalez}, {Hosseinzadeh}, {Huber}, {Fraser}, {Inserra}, {Kankare}, {Mattila}, {Magnier}, {Maguire}, {Lowe}, {Sollerman}, {Sullivan}, {Young}, \& {Valenti}}]{Anderson18}
{Anderson}, J.~P., {Dessart}, L., {Guti{\'e}rrez}, C.~P., {et~al.} 2018, Nature Astronomy, 2, 574, \dodoi{10.1038/s41550-018-0458-4}

\bibitem[{{Arcavi}(2018)}]{Arcavi18}
{Arcavi}, I. 2018, \apjl, 855, L23, \dodoi{10.3847/2041-8213/aab267}

\bibitem[{{Arcavi} {et~al.}(2010){Arcavi}, {Gal-Yam}, {Kasliwal}, {Quimby}, {Ofek}, {Kulkarni}, {Nugent}, {Cenko}, {Bloom}, {Sullivan}, {Howell}, {Poznanski}, {Filippenko}, {Law}, {Hook}, {J{\"o}nsson}, {Blake}, {Cooke}, {Dekany}, {Rahmer}, {Hale}, {Smith}, {Zolkower}, {Velur}, {Walters}, {Henning}, {Bui}, {McKenna}, \& {Jacobsen}}]{Arcavi10}
{Arcavi}, I., {Gal-Yam}, A., {Kasliwal}, M.~M., {et~al.} 2010, \apj, 721, 777, \dodoi{10.1088/0004-637X/721/1/777}

\bibitem[{{Asplund} {et~al.}(2009){Asplund}, {Grevesse}, {Sauval}, \& {Scott}}]{Asplund09}
{Asplund}, M., {Grevesse}, N., {Sauval}, A.~J., \& {Scott}, P. 2009, \araa, 47, 481, \dodoi{10.1146/annurev.astro.46.060407.145222}

\bibitem[{{Astropy Collaboration} {et~al.}(2013){Astropy Collaboration}, {Robitaille}, {Tollerud}, {Greenfield}, {Droettboom}, {Bray}, {Aldcroft}, {Davis}, {Ginsburg}, {Price-Whelan}, {Kerzendorf}, {Conley}, {Crighton}, {Barbary}, {Muna}, {Ferguson}, {Grollier}, {Parikh}, {Nair}, {Unther}, {Deil}, {Woillez}, {Conseil}, {Kramer}, {Turner}, {Singer}, {Fox}, {Weaver}, {Zabalza}, {Edwards}, {Azalee Bostroem}, {Burke}, {Casey}, {Crawford}, {Dencheva}, {Ely}, {Jenness}, {Labrie}, {Lim}, {Pierfederici}, {Pontzen}, {Ptak}, {Refsdal}, {Servillat}, \& {Streicher}}]{astropy}
{Astropy Collaboration}, {Robitaille}, T.~P., {Tollerud}, E.~J., {et~al.} 2013, \aap, 558, A33, \dodoi{10.1051/0004-6361/201322068}

\bibitem[{{Astropy Collaboration} {et~al.}(2018){Astropy Collaboration}, {Price-Whelan}, {Sip{\H{o}}cz}, {G{\"u}nther}, {Lim}, {Crawford}, {Conseil}, {Shupe}, {Craig}, {Dencheva}, {Ginsburg}, {VanderPlas}, {Bradley}, {P{\'e}rez-Su{\'a}rez}, {de Val-Borro}, {Aldcroft}, {Cruz}, {Robitaille}, {Tollerud}, {Ardelean}, {Babej}, {Bach}, {Bachetti}, {Bakanov}, {Bamford}, {Barentsen}, {Barmby}, {Baumbach}, {Berry}, {Biscani}, {Boquien}, {Bostroem}, {Bouma}, {Brammer}, {Bray}, {Breytenbach}, {Buddelmeijer}, {Burke}, {Calderone}, {Cano Rodr{\'\i}guez}, {Cara}, {Cardoso}, {Cheedella}, {Copin}, {Corrales}, {Crichton}, {D'Avella}, {Deil}, {Depagne}, {Dietrich}, {Donath}, {Droettboom}, {Earl}, {Erben}, {Fabbro}, {Ferreira}, {Finethy}, {Fox}, {Garrison}, {Gibbons}, {Goldstein}, {Gommers}, {Greco}, {Greenfield}, {Groener}, {Grollier}, {Hagen}, {Hirst}, {Homeier}, {Horton}, {Hosseinzadeh}, {Hu}, {Hunkeler}, {Ivezi{\'c}}, {Jain}, {Jenness}, {Kanarek}, {Kendrew}, {Kern}, {Kerzendorf}, {Khvalko}, {King}, {Kirkby}, {Kulkarni},
  {Kumar}, {Lee}, {Lenz}, {Littlefair}, {Ma}, {Macleod}, {Mastropietro}, {McCully}, {Montagnac}, {Morris}, {Mueller}, {Mumford}, {Muna}, {Murphy}, {Nelson}, {Nguyen}, {Ninan}, {N{\"o}the}, {Ogaz}, {Oh}, {Parejko}, {Parley}, {Pascual}, {Patil}, {Patil}, {Plunkett}, {Prochaska}, {Rastogi}, {Reddy Janga}, {Sabater}, {Sakurikar}, {Seifert}, {Sherbert}, {Sherwood-Taylor}, {Shih}, {Sick}, {Silbiger}, {Singanamalla}, {Singer}, {Sladen}, {Sooley}, {Sornarajah}, {Streicher}, {Teuben}, {Thomas}, {Tremblay}, {Turner}, {Terr{\'o}n}, {van Kerkwijk}, {de la Vega}, {Watkins}, {Weaver}, {Whitmore}, {Woillez}, {Zabalza}, \& {Astropy Contributors}}]{astropy2}
{Astropy Collaboration}, {Price-Whelan}, A.~M., {Sip{\H{o}}cz}, B.~M., {et~al.} 2018, \aj, 156, 123, \dodoi{10.3847/1538-3881/aabc4f}

\bibitem[{{Astropy Collaboration} {et~al.}(2022){Astropy Collaboration}, {Price-Whelan}, {Lim}, {Earl}, {Starkman}, {Bradley}, {Shupe}, {Patil}, {Corrales}, {Brasseur}, {N{\"o}the}, {Donath}, {Tollerud}, {Morris}, {Ginsburg}, {Vaher}, {Weaver}, {Tocknell}, {Jamieson}, {van Kerkwijk}, {Robitaille}, {Merry}, {Bachetti}, {G{\"u}nther}, {Aldcroft}, {Alvarado-Montes}, {Archibald}, {B{\'o}di}, {Bapat}, {Barentsen}, {Baz{\'a}n}, {Biswas}, {Boquien}, {Burke}, {Cara}, {Cara}, {Conroy}, {Conseil}, {Craig}, {Cross}, {Cruz}, {D'Eugenio}, {Dencheva}, {Devillepoix}, {Dietrich}, {Eigenbrot}, {Erben}, {Ferreira}, {Foreman-Mackey}, {Fox}, {Freij}, {Garg}, {Geda}, {Glattly}, {Gondhalekar}, {Gordon}, {Grant}, {Greenfield}, {Groener}, {Guest}, {Gurovich}, {Handberg}, {Hart}, {Hatfield-Dodds}, {Homeier}, {Hosseinzadeh}, {Jenness}, {Jones}, {Joseph}, {Kalmbach}, {Karamehmetoglu}, {Ka{\l}uszy{\'n}ski}, {Kelley}, {Kern}, {Kerzendorf}, {Koch}, {Kulumani}, {Lee}, {Ly}, {Ma}, {MacBride}, {Maljaars}, {Muna}, {Murphy}, {Norman},
  {O'Steen}, {Oman}, {Pacifici}, {Pascual}, {Pascual-Granado}, {Patil}, {Perren}, {Pickering}, {Rastogi}, {Roulston}, {Ryan}, {Rykoff}, {Sabater}, {Sakurikar}, {Salgado}, {Sanghi}, {Saunders}, {Savchenko}, {Schwardt}, {Seifert-Eckert}, {Shih}, {Jain}, {Shukla}, {Sick}, {Simpson}, {Singanamalla}, {Singer}, {Singhal}, {Sinha}, {Sip{\H{o}}cz}, {Spitler}, {Stansby}, {Streicher}, {{\v{S}}umak}, {Swinbank}, {Taranu}, {Tewary}, {Tremblay}, {de Val-Borro}, {Van Kooten}, {Vasovi{\'c}}, {Verma}, {de Miranda Cardoso}, {Williams}, {Wilson}, {Winkel}, {Wood-Vasey}, {Xue}, {Yoachim}, {Zhang}, {Zonca}, \& {Astropy Project Contributors}}]{astropy3}
{Astropy Collaboration}, {Price-Whelan}, A.~M., {Lim}, P.~L., {et~al.} 2022, \apj, 935, 167, \dodoi{10.3847/1538-4357/ac7c74}

\bibitem[{{Barnes} {et~al.}(2018){Barnes}, {Duffell}, {Liu}, {Modjaz}, {Bianco}, {Kasen}, \& {MacFadyen}}]{Barnes+18}
{Barnes}, J., {Duffell}, P.~C., {Liu}, Y., {et~al.} 2018, \apj, 860, 38, \dodoi{10.3847/1538-4357/aabf84}

\bibitem[{Becker(2015)}]{becker_hotpants_2015}
Becker, A. 2015, {HOTPANTS}: {High} {Order} {Transform} of {PSF} {ANd} {Template} {Subtraction}

\bibitem[{{Belkin} {et~al.}(2024){Belkin}, {Gompertz}, {Kumar}, {Ackley}, {Galloway}, {Jim{\'e}nez-Ibarra}, {Killestein}, {O'Neill}, {Wiersema}, {Malesani}, {Levan}, {Lyman}, {Dyer}, {Ulaczyk}, {Steeghs}, {Dhillon}, {O'Brien}, {Ramsay}, {Noysena}, {Kotak}, {Breton}, {Nuttall}, {Pall{\'e}}, {Pollacco}, {Awiphan}, {Burhanudin}, {Chote}, {Chrimes}, {Daw}, {Duffy}, {Eyles-Ferris}, {Godson}, {Heikkil{\"a}}, {Irawati}, {Kelsey}, {Kennedy}, {Littlefair}, {Makrygianni}, {Marsh}, {Mata S{\'a}nchez}, {Mattila}, {Maund}, {McCormac}, {Mkrtichian}, {Mullaney}, {Patel}, {Rana}, {Rol}, {Sawangwit}, {Stanway}, {Starling}, {Str{\o}m}, \& {Warwick}}]{Belkin24}
{Belkin}, S., {Gompertz}, B.~P., {Kumar}, A., {et~al.} 2024, Research Notes of the American Astronomical Society, 8, 6, \dodoi{10.3847/2515-5172/ad1876}

\bibitem[{{Branch} {et~al.}(2002){Branch}, {Benetti}, {Kasen}, {Baron}, {Jeffery}, {Hatano}, {Stathakis}, {Filippenko}, {Matheson}, {Pastorello}, {Altavilla}, {Cappellaro}, {Rizzi}, {Turatto}, {Li}, {Leonard}, \& {Shields}}]{Branch02}
{Branch}, D., {Benetti}, S., {Kasen}, D., {et~al.} 2002, \apj, 566, 1005, \dodoi{10.1086/338127}

\bibitem[{{Bunker} {et~al.}(2023){Bunker}, {Cameron}, {Curtis-Lake}, {Jakobsen}, {Carniani}, {Curti}, {Witstok}, {Maiolino}, {D'Eugenio}, {Looser}, {Willott}, {Bonaventura}, {Hainline}, {Uebler}, {Willmer}, {Saxena}, {Smit}, {Alberts}, {Arribas}, {Baker}, {Baum}, {Bhatawdekar}, {Bowler}, {Boyett}, {Charlot}, {Chen}, {Chevallard}, {Circosta}, {DeCoursey}, {de Graaff}, {Egami}, {Eisenstein}, {Endsley}, {Ferruit}, {Giardino}, {Hausen}, {Helton}, {Hviding}, {Ji}, {Johnson}, {Jones}, {Kumari}, {Laseter}, {Luetzgendorf}, {Maseda}, {Nelson}, {Parlanti}, {Perna}, {Rawle}, {Rix}, {Rieke}, {Robertson}, {Rodriguez Del Pino}, {Sandles}, {Scholtz}, {Sharpe}, {Skarbinski}, {Stark}, {Sun}, {Tacchella}, {Topping}, {Villanueva}, {Wallace}, {Williams}, \& {Woodrum}}]{Bunker23}
{Bunker}, A.~J., {Cameron}, A.~J., {Curtis-Lake}, E., {et~al.} 2023, arXiv e-prints, arXiv:2306.02467, \dodoi{10.48550/arXiv.2306.02467}

\bibitem[{Bushouse {et~al.}(2024)Bushouse, Eisenhamer, Dencheva, Davies, Greenfield, Morrison, Hodge, Simon, Grumm, Droettboom, Slavich, Sosey, Pauly, Miller, Jedrzejewski, Hack, Davis, Crawford, Law, Gordon, Regan, Cara, MacDonald, Bradley, Shanahan, Jamieson, Teodoro, \& Williams}]{Bushouse_JWST_Calibration_Pipeline_2024}
Bushouse, H., Eisenhamer, J., Dencheva, N., {et~al.} 2024, {JWST Calibration Pipeline}, 1.14.0, \dodoi{10.5281/zenodo.7038885}

\bibitem[{{Cano}(2013)}]{Cano13}
{Cano}, Z. 2013, \mnras, 434, 1098, \dodoi{10.1093/mnras/stt1048}

\bibitem[{{Cao} {et~al.}(2013){Cao}, {Kasliwal}, {Arcavi}, {Horesh}, {Hancock}, {Valenti}, {Cenko}, {Kulkarni}, {Gal-Yam}, {Gorbikov}, {Ofek}, {Sand}, {Yaron}, {Graham}, {Silverman}, {Wheeler}, {Marion}, {Walker}, {Mazzali}, {Howell}, {Li}, {Kong}, {Bloom}, {Nugent}, {Surace}, {Masci}, {Carpenter}, {Degenaar}, \& {Gelino}}]{cao13}
{Cao}, Y., {Kasliwal}, M.~M., {Arcavi}, I., {et~al.} 2013, \apjl, 775, L7, \dodoi{10.1088/2041-8205/775/1/L7}

\bibitem[{{Carnall} {et~al.}(2019){Carnall}, {Leja}, {Johnson}, {McLure}, {Dunlop}, \& {Conroy}}]{Carnall19}
{Carnall}, A.~C., {Leja}, J., {Johnson}, B.~D., {et~al.} 2019, \apj, 873, 44, \dodoi{10.3847/1538-4357/ab04a2}

\bibitem[{{Carnall} {et~al.}(2018){Carnall}, {McLure}, {Dunlop}, \& {Dav{\'e}}}]{Carnall18}
{Carnall}, A.~C., {McLure}, R.~J., {Dunlop}, J.~S., \& {Dav{\'e}}, R. 2018, \mnras, 480, 4379, \dodoi{10.1093/mnras/sty2169}

\bibitem[{{Clocchiatti} {et~al.}(1996){Clocchiatti}, {Wheeler}, {Benetti}, \& {Frueh}}]{Clocchiatti+96}
{Clocchiatti}, A., {Wheeler}, J.~C., {Benetti}, S., \& {Frueh}, M. 1996, \apj, 459, 547, \dodoi{10.1086/176919}

\bibitem[{{Corsi} {et~al.}(2012){Corsi}, {Ofek}, {Gal-Yam}, {Frail}, {Poznanski}, {Mazzali}, {Kulkarni}, {Kasliwal}, {Arcavi}, {Ben-Ami}, {Cenko}, {Filippenko}, {Fox}, {Horesh}, {Howell}, {Kleiser}, {Nakar}, {Rabinak}, {Sari}, {Silverman}, {Xu}, {Bloom}, {Law}, {Nugent}, \& {Quimby}}]{Corsi+12}
{Corsi}, A., {Ofek}, E.~O., {Gal-Yam}, A., {et~al.} 2012, \apjl, 747, L5, \dodoi{10.1088/2041-8205/747/1/L5}

\bibitem[{{Curti} {et~al.}(2023){Curti}, {D'Eugenio}, {Carniani}, {Maiolino}, {Sandles}, {Witstok}, {Baker}, {Bennett}, {Piotrowska}, {Tacchella}, {Charlot}, {Nakajima}, {Maheson}, {Mannucci}, {Amiri}, {Arribas}, {Belfiore}, {Bonaventura}, {Bunker}, {Chevallard}, {Cresci}, {Curtis-Lake}, {Hayden-Pawson}, {Jones}, {Kumari}, {Laseter}, {Looser}, {Marconi}, {Maseda}, {Scholtz}, {Smit}, {{\"U}bler}, \& {Wallace}}]{Curti23}
{Curti}, M., {D'Eugenio}, F., {Carniani}, S., {et~al.} 2023, \mnras, 518, 425, \dodoi{10.1093/mnras/stac2737}

\bibitem[{{De} {et~al.}(2019){De}, {Kasliwal}, {Polin}, {Nugent}, {Bildsten}, {Adams}, {Bellm}, {Blagorodnova}, {Burdge}, {Cannella}, {Cenko}, {Dekany}, {Feeney}, {Hale}, {Fremling}, {Graham}, {Ho}, {Jencson}, {Kulkarni}, {Laher}, {Masci}, {Miller}, {Patterson}, {Rebbapragada}, {Riddle}, {Shupe}, \& {Smith}}]{De19}
{De}, K., {Kasliwal}, M.~M., {Polin}, A., {et~al.} 2019, \apjl, 873, L18, \dodoi{10.3847/2041-8213/ab0aec}

\bibitem[{{DeCoursey} {et~al.}(2024){DeCoursey}, {Egami}, {Pierel}, {Sun}, {Rest}, {Coulter}, {Engesser}, {Siebert}, {Hainline}, {Johnson}, {Bunker}, {Cargile}, {Charlot}, {Chen}, {Curti}, {DeFour-Remy}, {Eisenstein}, {Fox}, {Gezari}, {Gomez}, {Jencson}, {Joshi}, {Khairnar}, {Lyu}, {Maiolino}, {Moriya}, {Quimby}, {Rieke}, {Rieke}, {Robertson}, {Shahbandeh}, {Strolger}, {Tacchella}, {Wang}, {Williams}, {Willmer}, {Willott}, \& {Zenati}}]{Decoursey24}
{DeCoursey}, C., {Egami}, E., {Pierel}, J. D.~R., {et~al.} 2024, arXiv e-prints, arXiv:2406.05060.
\newblock \doarXiv{2406.05060}

\bibitem[{{Dessart} {et~al.}(2012){Dessart}, {Hillier}, {Li}, \& {Woosley}}]{Dessart+12}
{Dessart}, L., {Hillier}, D.~J., {Li}, C., \& {Woosley}, S. 2012, \mnras, 424, 2139, \dodoi{10.1111/j.1365-2966.2012.21374.x}

\bibitem[{{Dessart} {et~al.}(2020){Dessart}, {Yoon}, {Aguilera-Dena}, \& {Langer}}]{Dessart+20}
{Dessart}, L., {Yoon}, S.-C., {Aguilera-Dena}, D.~R., \& {Langer}, N. 2020, \aap, 642, A106, \dodoi{10.1051/0004-6361/202038763}

\bibitem[{{Drout} {et~al.}(2011){Drout}, {Soderberg}, {Gal-Yam}, {Cenko}, {Fox}, {Leonard}, {Sand}, {Moon}, {Arcavi}, \& {Green}}]{Drout11}
{Drout}, M.~R., {Soderberg}, A.~M., {Gal-Yam}, A., {et~al.} 2011, \apj, 741, 97, \dodoi{10.1088/0004-637X/741/2/97}

\bibitem[{{Drout} {et~al.}(2016){Drout}, {Milisavljevic}, {Parrent}, {Margutti}, {Kamble}, {Soderberg}, {Challis}, {Chornock}, {Fong}, {Frank}, {Gehrels}, {Graham}, {Hsiao}, {Itagaki}, {Kasliwal}, {Kirshner}, {Macomb}, {Marion}, {Norris}, \& {Phillips}}]{Drout16}
{Drout}, M.~R., {Milisavljevic}, D., {Parrent}, J., {et~al.} 2016, \apj, 821, 57, \dodoi{10.3847/0004-637X/821/1/57}

\bibitem[{{Eisenstein} {et~al.}(2011){Eisenstein}, {Weinberg}, {Agol}, {Aihara}, {Allende Prieto}, {Anderson}, {Arns}, {Aubourg}, {Bailey}, {Balbinot}, {Barkhouser}, {Beers}, {Berlind}, {Bickerton}, {Bizyaev}, {Blanton}, {Bochanski}, {Bolton}, {Bosman}, {Bovy}, {Brandt}, {Breslauer}, {Brewington}, {Brinkmann}, {Brown}, {Brownstein}, {Burger}, {Busca}, {Campbell}, {Cargile}, {Carithers}, {Carlberg}, {Carr}, {Chang}, {Chen}, {Chiappini}, {Comparat}, {Connolly}, {Cortes}, {Croft}, {Cunha}, {da Costa}, {Davenport}, {Dawson}, {De Lee}, {Porto de Mello}, {de Simoni}, {Dean}, {Dhital}, {Ealet}, {Ebelke}, {Edmondson}, {Eiting}, {Escoffier}, {Esposito}, {Evans}, {Fan}, {Femen{\'\i}a Castell{\'a}}, {Dutra Ferreira}, {Fitzgerald}, {Fleming}, {Font-Ribera}, {Ford}, {Frinchaboy}, {Garc{\'\i}a P{\'e}rez}, {Gaudi}, {Ge}, {Ghezzi}, {Gillespie}, {Gilmore}, {Girardi}, {Gott}, {Gould}, {Grebel}, {Gunn}, {Hamilton}, {Harding}, {Harris}, {Hawley}, {Hearty}, {Hennawi}, {Gonz{\'a}lez Hern{\'a}ndez}, {Ho}, {Hogg}, {Holtzman},
  {Honscheid}, {Inada}, {Ivans}, {Jiang}, {Jiang}, {Johnson}, {Jordan}, {Jordan}, {Kauffmann}, {Kazin}, {Kirkby}, {Klaene}, {Knapp}, {Kneib}, {Kochanek}, {Koesterke}, {Kollmeier}, {Kron}, {Lampeitl}, {Lang}, {Lawler}, {Le Goff}, {Lee}, {Lee}, {Leisenring}, {Lin}, {Liu}, {Long}, {Loomis}, {Lucatello}, {Lundgren}, {Lupton}, {Ma}, {Ma}, {MacDonald}, {Mack}, {Mahadevan}, {Maia}, {Majewski}, {Makler}, {Malanushenko}, {Malanushenko}, {Mandelbaum}, {Maraston}, {Margala}, {Maseman}, {Masters}, {McBride}, {McDonald}, {McGreer}, {McMahon}, {Mena Requejo}, {M{\'e}nard}, {Miralda-Escud{\'e}}, {Morrison}, {Mullally}, {Muna}, {Murayama}, {Myers}, {Naugle}, {Neto}, {Nguyen}, {Nichol}, {Nidever}, {O'Connell}, {Ogando}, {Olmstead}, {Oravetz}, {Padmanabhan}, {Paegert}, {Palanque-Delabrouille}, {Pan}, {Pandey}, {Parejko}, {P{\^a}ris}, {Pellegrini}, {Pepper}, {Percival}, {Petitjean}, {Pfaffenberger}, {Pforr}, {Phleps}, {Pichon}, {Pieri}, {Prada}, {Price-Whelan}, {Raddick}, {Ramos}, {Reid}, {Reyle}, {Rich}, {Richards}, {Rieke},
  {Rieke}, {Rix}, {Robin}, {Rocha-Pinto}, {Rockosi}, {Roe}, {Rollinde}, {Ross}, {Ross}, {Rossetto}, {S{\'a}nchez}, {Santiago}, {Sayres}, {Schiavon}, {Schlegel}, {Schlesinger}, {Schmidt}, {Schneider}, {Sellgren}, {Shelden}, {Sheldon}, {Shetrone}, {Shu}, {Silverman}, {Simmerer}, {Simmons}, {Sivarani}, {Skrutskie}, {Slosar}, {Smee}, {Smith}, {Snedden}, {Stassun}, {Steele}, {Steinmetz}, {Stockett}, {Stollberg}, {Strauss}, {Szalay}, {Tanaka}, {Thakar}, {Thomas}, {Tinker}, {Tofflemire}, {Tojeiro}, {Tremonti}, {Vargas Maga{\~n}a}, {Verde}, {Vogt}, {Wake}, {Wan}, {Wang}, {Weaver}, {White}, {White}, {Wilson}, {Wisniewski}, {Wood-Vasey}, {Yanny}, {Yasuda}, {Y{\`e}che}, {York}, {Young}, {Zasowski}, {Zehavi}, \& {Zhao}}]{Eisenstein11}
{Eisenstein}, D.~J., {Weinberg}, D.~H., {Agol}, E., {et~al.} 2011, \aj, 142, 72, \dodoi{10.1088/0004-6256/142/3/72}

\bibitem[{{Eisenstein} {et~al.}(2023){Eisenstein}, {Willott}, {Alberts}, {Arribas}, {Bonaventura}, {Bunker}, {Cameron}, {Carniani}, {Charlot}, {Curtis-Lake}, {D'Eugenio}, {Endsley}, {Ferruit}, {Giardino}, {Hainline}, {Hausen}, {Jakobsen}, {Johnson}, {Maiolino}, {Rieke}, {Rieke}, {Rix}, {Robertson}, {Stark}, {Tacchella}, {Williams}, {Willmer}, {Baker}, {Baum}, {Bhatawdekar}, {Boyett}, {Chen}, {Chevallard}, {Circosta}, {Curti}, {Danhaive}, {DeCoursey}, {de Graaff}, {Dressler}, {Egami}, {Helton}, {Hviding}, {Ji}, {Jones}, {Kumari}, {L{\"u}tzgendorf}, {Laseter}, {Looser}, {Lyu}, {Maseda}, {Nelson}, {Parlanti}, {Perna}, {Pusk{\'a}s}, {Rawle}, {Rodr{\'\i}guez Del Pino}, {Sandles}, {Saxena}, {Scholtz}, {Sharpe}, {Shivaei}, {Silcock}, {Simmonds}, {Skarbinski}, {Smit}, {Stone}, {Suess}, {Sun}, {Tang}, {Topping}, {{\"U}bler}, {Villanueva}, {Wallace}, {Whitler}, {Witstok}, \& {Woodrum}}]{Eisenstein23}
{Eisenstein}, D.~J., {Willott}, C., {Alberts}, S., {et~al.} 2023, arXiv e-prints, arXiv:2306.02465, \dodoi{10.48550/arXiv.2306.02465}

\bibitem[{{Eldridge} \& {Maund}(2016)}]{eldridge16}
{Eldridge}, J.~J., \& {Maund}, J.~R. 2016, \mnras, 461, L117, \dodoi{10.1093/mnrasl/slw099}

\bibitem[{{Elias-Rosa} {et~al.}(2013){Elias-Rosa}, {Pastorello}, {Maund}, {Takats}, {Fraser}, {Smartt}, {Benetti}, {Pignata}, {Sand}, \& {Valenti}}]{EliasRosa13}
{Elias-Rosa}, N., {Pastorello}, A., {Maund}, J.~R., {et~al.} 2013, \mnras, 436, L109, \dodoi{10.1093/mnrasl/slt124}

\bibitem[{{Ferruit} {et~al.}(2022){Ferruit}, {Jakobsen}, {Giardino}, {Rawle}, {Alves de Oliveira}, {Arribas}, {Beck}, {Birkmann}, {B{\"o}ker}, {Bunker}, {Charlot}, {de Marchi}, {Franx}, {Henry}, {Karakla}, {Kassin}, {Kumari}, {L{\'o}pez-Caniego}, {L{\"u}tzgendorf}, {Maiolino}, {Manjavacas}, {Marston}, {Moseley}, {Muzerolle}, {Pirzkal}, {Rauscher}, {Rix}, {Sabbi}, {Sirianni}, {te Plate}, {Valenti}, {Willott}, \& {Zeidler}}]{Ferruit22}
{Ferruit}, P., {Jakobsen}, P., {Giardino}, G., {et~al.} 2022, \aap, 661, A81, \dodoi{10.1051/0004-6361/202142673}

\bibitem[{{Filippenko} {et~al.}(1995){Filippenko}, {Barth}, {Bower}, {Ho}, {Stringfellow}, {Goodrich}, \& {Porter}}]{Filippenko95}
{Filippenko}, A.~V., {Barth}, A.~J., {Bower}, G.~C., {et~al.} 1995, \aj, 110, 2261, \dodoi{10.1086/117687}

\bibitem[{{Folatelli} {et~al.}(2016){Folatelli}, {Van Dyk}, {Kuncarayakti}, {Maeda}, {Bersten}, {Nomoto}, {Pignata}, {Hamuy}, {Quimby}, {Zheng}, {Filippenko}, {Clubb}, {Smith}, {Elias-Rosa}, {Foley}, \& {Miller}}]{folatelli16}
{Folatelli}, G., {Van Dyk}, S.~D., {Kuncarayakti}, H., {et~al.} 2016, \apjl, 825, L22, \dodoi{10.3847/2041-8205/825/2/L22}

\bibitem[{{Foley} {et~al.}(2003){Foley}, {Papenkova}, {Swift}, {Filippenko}, {Li}, {Mazzali}, {Chornock}, {Leonard}, \& {Van Dyk}}]{Foley03}
{Foley}, R.~J., {Papenkova}, M.~S., {Swift}, B.~J., {et~al.} 2003, \pasp, 115, 1220

\bibitem[{{Fox} {et~al.}(2022){Fox}, {Van Dyk}, {Williams}, {Drout}, {Zapartas}, {Smith}, {Milisavljevic}, {Andrews}, {Bostroem}, {Filippenko}, {Gomez}, {Kelly}, {de Mink}, {Pierel}, {Rest}, {Ryder}, {Sravan}, {Strolger}, {Wang}, \& {Weil}}]{fox22}
{Fox}, O.~D., {Van Dyk}, S.~D., {Williams}, B.~F., {et~al.} 2022, \apjl, 929, L15, \dodoi{10.3847/2041-8213/ac5890}

\bibitem[{{Gagliano} {et~al.}(2022){Gagliano}, {Izzo}, {Kilpatrick}, {Mockler}, {Jacobson-Gal{\'a}n}, {Terreran}, {Dimitriadis}, {Zenati}, {Auchettl}, {Drout}, {Narayan}, {Foley}, {Margutti}, {Rest}, {Jones}, {Aganze}, {Aleo}, {Burgasser}, {Coulter}, {Gerasimov}, {Gall}, {Hjorth}, {Hsu}, {Magnier}, {Mandel}, {Piro}, {Rojas-Bravo}, {Siebert}, {Stacey}, {Stroh}, {Swift}, {Taggart}, {Tinyanont}, \& {Tinyanont}}]{gagliano22}
{Gagliano}, A., {Izzo}, L., {Kilpatrick}, C.~D., {et~al.} 2022, \apj, 924, 55, \dodoi{10.3847/1538-4357/ac35ec}

\bibitem[{{Gal-Yam}(2017)}]{Gal-Yam17}
{Gal-Yam}, A. 2017, in Handbook of Supernovae, ed. A.~W. {Alsabti} \& P.~{Murdin}, 195, \dodoi{10.1007/978-3-319-21846-5_35}

\bibitem[{{Galama} {et~al.}(1998){Galama}, {Vreeswijk}, {van Paradijs}, {Kouveliotou}, {Augusteijn}, {B{\"o}hnhardt}, {Brewer}, {Doublier}, {Gonzalez}, {Leibundgut}, {Lidman}, {Hainaut}, {Patat}, {Heise}, {in't Zand}, {Hurley}, {Groot}, {Strom}, {Mazzali}, {Iwamoto}, {Nomoto}, {Umeda}, {Nakamura}, {Young}, {Suzuki}, {Shigeyama}, {Koshut}, {Kippen}, {Robinson}, {de Wildt}, {Wijers}, {Tanvir}, {Greiner}, {Pian}, {Palazzi}, {Frontera}, {Masetti}, {Nicastro}, {Feroci}, {Costa}, {Piro}, {Peterson}, {Tinney}, {Boyle}, {Cannon}, {Stathakis}, {Sadler}, {Begam}, \& {Ianna}}]{Galama+98Natur}
{Galama}, T.~J., {Vreeswijk}, P.~M., {van Paradijs}, J., {et~al.} 1998, \nat, 395, 670, \dodoi{10.1038/27150}

\bibitem[{{Georgy} {et~al.}(2009){Georgy}, {Meynet}, {Walder}, {Folini}, \& {Maeder}}]{Georgy09}
{Georgy}, C., {Meynet}, G., {Walder}, R., {Folini}, D., \& {Maeder}, A. 2009, \aap, 502, 611, \dodoi{10.1051/0004-6361/200811339}

\bibitem[{{Goldwasser} {et~al.}(2022){Goldwasser}, {Yaron}, {Sass}, {Irani}, {Gal-Yam}, \& {Howell}}]{Goldwasser22}
{Goldwasser}, S., {Yaron}, O., {Sass}, A., {et~al.} 2022, Transient Name Server AstroNote, 191, 1

\bibitem[{{Guti{\'e}rrez} {et~al.}(2018){Guti{\'e}rrez}, {Anderson}, {Sullivan}, {Dessart}, {Gonz{\'a}lez-Gaitan}, {Galbany}, {Dimitriadis}, {Arcavi}, {Bufano}, {Chen}, {Dennefeld}, {Gromadzki}, {Haislip}, {Hosseinzadeh}, {Howell}, {Inserra}, {Kankare}, {Leloudas}, {Maguire}, {McCully}, {Morrell}, {Olivares E}, {Pignata}, {Reichart}, {Reynolds}, {Smartt}, {Sollerman}, {Taddia}, {Tak{\'a}ts}, {Terreran}, {Valenti}, \& {Young}}]{Gutierrez18}
{Guti{\'e}rrez}, C.~P., {Anderson}, J.~P., {Sullivan}, M., {et~al.} 2018, \mnras, 479, 3232, \dodoi{10.1093/mnras/sty1581}

\bibitem[{{Howell} {et~al.}(2005){Howell}, {Sullivan}, {Perrett}, {Bronder}, {Hook}, {Astier}, {Aubourg}, {Balam}, {Basa}, {Carlberg}, {Fabbro}, {Fouchez}, {Guy}, {Lafoux}, {Neill}, {Pain}, {Palanque-Delabrouille}, {Pritchet}, {Regnault}, {Rich}, {Taillet}, {Knop}, {McMahon}, {Perlmutter}, \& {Walton}}]{Howell05}
{Howell}, D.~A., {Sullivan}, M., {Perrett}, K., {et~al.} 2005, \apj, 634, 1190, \dodoi{10.1086/497119}

\bibitem[{{Iwamoto} {et~al.}(1999){Iwamoto}, {Brachwitz}, {Nomoto}, {Kishimoto}, {Umeda}, {Hix}, \& {Thielemann}}]{Iwamoto99}
{Iwamoto}, K., {Brachwitz}, F., {Nomoto}, K., {et~al.} 1999, \apjs, 125, 439, \dodoi{10.1086/313278}

\bibitem[{{Jacobson-Gal{\'a}n} {et~al.}(2020){Jacobson-Gal{\'a}n}, {Polin}, {Foley}, {Dimitriadis}, {Kilpatrick}, {Margutti}, {Coulter}, {Jha}, {Jones}, {Kirshner}, {Pan}, {Piro}, {Rest}, \& {Rojas-Bravo}}]{Jacobson-Galan20}
{Jacobson-Gal{\'a}n}, W.~V., {Polin}, A., {Foley}, R.~J., {et~al.} 2020, \apj, 896, 165, \dodoi{10.3847/1538-4357/ab94b8}

\bibitem[{{Jakobsen} {et~al.}(2022){Jakobsen}, {Ferruit}, {Alves de Oliveira}, {Arribas}, {Bagnasco}, {Barho}, {Beck}, {Birkmann}, {B{\"o}ker}, {Bunker}, {Charlot}, {de Jong}, {de Marchi}, {Ehrenwinkler}, {Falcolini}, {Fels}, {Franx}, {Franz}, {Funke}, {Giardino}, {Gnata}, {Holota}, {Honnen}, {Jensen}, {Jentsch}, {Johnson}, {Jollet}, {Karl}, {Kling}, {K{\"o}hler}, {Kolm}, {Kumari}, {Lander}, {Lemke}, {L{\'o}pez-Caniego}, {L{\"u}tzgendorf}, {Maiolino}, {Manjavacas}, {Marston}, {Maschmann}, {Maurer}, {Messerschmidt}, {Moseley}, {Mosner}, {Mott}, {Muzerolle}, {Pirzkal}, {Pittet}, {Plitzke}, {Posselt}, {Rapp}, {Rauscher}, {Rawle}, {Rix}, {R{\"o}del}, {Rumler}, {Sabbi}, {Salvignol}, {Schmid}, {Sirianni}, {Smith}, {Strada}, {te Plate}, {Valenti}, {Wettemann}, {Wiehe}, {Wiesmayer}, {Willott}, {Wright}, {Zeidler}, \& {Zincke}}]{Jakobsen2022}
{Jakobsen}, P., {Ferruit}, P., {Alves de Oliveira}, C., {et~al.} 2022, \aap, 661, A80, \dodoi{10.1051/0004-6361/202142663}

\bibitem[{{Janka} {et~al.}(2016){Janka}, {Melson}, \& {Summa}}]{Janka16}
{Janka}, H.-T., {Melson}, T., \& {Summa}, A. 2016, Annual Review of Nuclear and Particle Science, 66, 341, \dodoi{10.1146/annurev-nucl-102115-044747}

\bibitem[{{Jdadf Developers} {et~al.}(2023){Jdadf Developers}, {Averbukh}, {Bradley}, {Buikhuizen}, {Busko}, {Cherinka}, {Conroy}, {Earl}, {Fox}, {Geda}, {Jones}, {Karatay}, {Kotler}, {Lim}, {Morris}, {Nguyen}, {O'\#03teen}, {Ogaz}, {Ogle}, {Otor}, {Pacifici}, {Robitaille}, {Shanahan}, {Tollerud}, \& {Volfman}}]{Jdaviz}
{Jdadf Developers}, {Averbukh}, J., {Bradley}, L., {et~al.} 2023, {Jdaviz: JWST astronomical data analysis tools in the Jupyter platform}, Astrophysics Source Code Library, record ascl:2307.001

\bibitem[{{Kelly} \& {Kirshner}(2012)}]{Kelly12}
{Kelly}, P.~L., \& {Kirshner}, R.~P. 2012, \apj, 759, 107, \dodoi{10.1088/0004-637X/759/2/107}

\bibitem[{{Kerzendorf} {et~al.}(2014){Kerzendorf}, {Taubenberger}, {Seitenzahl}, \& {Ruiter}}]{Kerzendorf14}
{Kerzendorf}, W.~E., {Taubenberger}, S., {Seitenzahl}, I.~R., \& {Ruiter}, A.~J. 2014, \apjl, 796, L26, \dodoi{10.1088/2041-8205/796/2/L26}

\bibitem[{{Kilpatrick} {et~al.}(2021){Kilpatrick}, {Drout}, {Auchettl}, {Dimitriadis}, {Foley}, {Jones}, {DeMarchi}, {French}, {Gall}, {Hjorth}, {Jacobson-Gal{\'a}n}, {Margutti}, {Piro}, {Ramirez-Ruiz}, {Rest}, \& {Rojas-Bravo}}]{kilpatrick21}
{Kilpatrick}, C.~D., {Drout}, M.~R., {Auchettl}, K., {et~al.} 2021, \mnras, 504, 2073, \dodoi{10.1093/mnras/stab838}

\bibitem[{{Kwok} {et~al.}(2022){Kwok}, {Williamson}, {Jha}, {Modjaz}, {Camacho-Neves}, {Foley}, {Garnavich}, {Maeda}, {Milisavljevic}, {Pandya}, {Dai}, {McCully}, {Pritchard}, \& {Singhal}}]{Kwok22}
{Kwok}, L.~A., {Williamson}, M., {Jha}, S.~W., {et~al.} 2022, \apj, 937, 40, \dodoi{10.3847/1538-4357/ac8989}

\bibitem[{{Li} {et~al.}(2011){Li}, {Leaman}, {Chornock}, {Filippenko}, {Poznanski}, {Ganeshalingam}, {Wang}, {Modjaz}, {Jha}, {Foley}, \& {Smith}}]{Li11:rate2}
{Li}, W., {Leaman}, J., {Chornock}, R., {et~al.} 2011, \mnras, 412, 1441, \dodoi{10.1111/j.1365-2966.2011.18160.x}

\bibitem[{{Lyman} {et~al.}(2016){Lyman}, {Bersier}, {James}, {Mazzali}, {Eldridge}, {Fraser}, \& {Pian}}]{Lyman16}
{Lyman}, J.~D., {Bersier}, D., {James}, P.~A., {et~al.} 2016, \mnras, 457, 328, \dodoi{10.1093/mnras/stv2983}

\bibitem[{{Maeda} {et~al.}(2003){Maeda}, {Mazzali}, {Deng}, {Nomoto}, {Yoshii}, {Tomita}, \& {Kobayashi}}]{Maeda03}
{Maeda}, K., {Mazzali}, P.~A., {Deng}, J., {et~al.} 2003, \apj, 593, 931, \dodoi{10.1086/376591}

\bibitem[{{Mannucci} {et~al.}(2011){Mannucci}, {Salvaterra}, \& {Campisi}}]{Mannucci11}
{Mannucci}, F., {Salvaterra}, R., \& {Campisi}, M.~A. 2011, \mnras, 414, 1263, \dodoi{10.1111/j.1365-2966.2011.18459.x}

\bibitem[{{Matheson} {et~al.}(2001){Matheson}, {Filippenko}, {Li}, {Leonard}, \& {Shields}}]{Matheson+01}
{Matheson}, T., {Filippenko}, A.~V., {Li}, W., {Leonard}, D.~C., \& {Shields}, J.~C. 2001, \aj, 121, 1648, \dodoi{10.1086/319390}

\bibitem[{{Mazzali} {et~al.}(2002){Mazzali}, {Deng}, {Maeda}, {Nomoto}, {Umeda}, {Hatano}, {Iwamoto}, {Yoshii}, {Kobayashi}, {Minezaki}, {Doi}, {Enya}, {Tomita}, {Smartt}, {Kinugasa}, {Kawakita}, {Ayani}, {Kawabata}, {Yamaoka}, {Qiu}, {Motohara}, {Gerardy}, {Fesen}, {Kawabata}, {Iye}, {Kashikawa}, {Kosugi}, {Ohyama}, {Takada-Hidai}, {Zhao}, {Chornock}, {Filippenko}, {Benetti}, \& {Turatto}}]{Mazzali02}
{Mazzali}, P.~A., {Deng}, J., {Maeda}, K., {et~al.} 2002, \apjl, 572, L61, \dodoi{10.1086/341504}

\bibitem[{{Milisavljevic} {et~al.}(2015){Milisavljevic}, {Margutti}, {Parrent}, {Soderberg}, {Fesen}, {Mazzali}, {Maeda}, {Sanders}, {Cenko}, {Silverman}, {Filippenko}, {Kamble}, {Chakraborti}, {Drout}, {Kirshner}, {Pickering}, {Kawabata}, {Hattori}, {Hsiao}, {Stritzinger}, {Marion}, {Vinko}, \& {Wheeler}}]{Milisavljevic15}
{Milisavljevic}, D., {Margutti}, R., {Parrent}, J.~T., {et~al.} 2015, \apj, 799, 51, \dodoi{10.1088/0004-637X/799/1/51}

\bibitem[{{Modjaz} {et~al.}(2016){Modjaz}, {Liu}, {Bianco}, \& {Graur}}]{Modjaz+16}
{Modjaz}, M., {Liu}, Y.~Q., {Bianco}, F.~B., \& {Graur}, O. 2016, \apj, 832, 108, \dodoi{10.3847/0004-637X/832/2/108}

\bibitem[{{Modjaz} {et~al.}(2006){Modjaz}, {Stanek}, {Garnavich}, {Berlind}, {Blondin}, {Brown}, {Calkins}, {Challis}, {Diamond-Stanic}, {Hao}, {Hicken}, {Kirshner}, \& {Prieto}}]{Modjaz06}
{Modjaz}, M., {Stanek}, K.~Z., {Garnavich}, P.~M., {et~al.} 2006, \apjl, 645, L21, \dodoi{10.1086/505906}

\bibitem[{{Modjaz} {et~al.}(2019){Modjaz}, {Bianco}, {Siwek}, {Huang}, {Perley}, {Fierroz}, {Liu}, {Arcavi}, {Gal-Yam}, {Blagorodnova}, {Cenko}, {Filippenko}, {Kasliwal}, {Kulkarni}, {Schulze}, {Taggart}, \& {Zhen}}]{Modjaz+19}
{Modjaz}, M., {Bianco}, F.~B., {Siwek}, M., {et~al.} 2019, arXiv e-prints, arXiv:1901.00872.
\newblock \doarXiv{1901.00872}

\bibitem[{{Modjaz} {et~al.}(2020){Modjaz}, {Bianco}, {Siwek}, {Huang}, {Perley}, {Fierroz}, {Liu}, {Arcavi}, {Gal-Yam}, {Filippenko}, {Blagorodnova}, {Cenko}, {Kasliwal}, {Kulkarni}, {Schulze}, {Taggart}, \& {Zheng}}]{Modjaz20}
---. 2020, \apj, 892, 153, \dodoi{10.3847/1538-4357/ab4185}

\bibitem[{{Morishita} {et~al.}(2024){Morishita}, {Stiavelli}, {Grillo}, {Rosati}, {Schuldt}, {Trenti}, {Bergamini}, {Boyett}, {Chary}, {Leethochawalit}, {Roberts-Borsani}, {Treu}, \& {Vanzella}}]{Morishita24}
{Morishita}, T., {Stiavelli}, M., {Grillo}, C., {et~al.} 2024, arXiv e-prints, arXiv:2402.14084, \dodoi{10.48550/arXiv.2402.14084}

\bibitem[{{Nakajima} {et~al.}(2023){Nakajima}, {Ouchi}, {Isobe}, {Harikane}, {Zhang}, {Ono}, {Umeda}, \& {Oguri}}]{Nakajima23}
{Nakajima}, K., {Ouchi}, M., {Isobe}, Y., {et~al.} 2023, \apjs, 269, 33, \dodoi{10.3847/1538-4365/acd556}

\bibitem[{{Pettini} \& {Pagel}(2004)}]{Pettini04}
{Pettini}, M., \& {Pagel}, B.~E.~J. 2004, \mnras, 348, L59, \dodoi{10.1111/j.1365-2966.2004.07591.x}

\bibitem[{{Pian} {et~al.}(2006){Pian}, {Mazzali}, {Masetti}, {Ferrero}, {Klose}, {Palazzi}, {Ramirez-Ruiz}, {Woosley}, {Kouveliotou}, {Deng}, {Filippenko}, {Foley}, {Fynbo}, {Kann}, {Li}, {Hjorth}, {Nomoto}, {Patat}, {Sauer}, {Sollerman}, {Vreeswijk}, {Guenther}, {Levan}, {O'Brien}, {Tanvir}, {Wijers}, {Dumas}, {Hainaut}, {Wong}, {Baade}, {Wang}, {Amati}, {Cappellaro}, {Castro-Tirado}, {Ellison}, {Frontera}, {Fruchter}, {Greiner}, {Kawabata}, {Ledoux}, {Maeda}, {M{\o}ller}, {Nicastro}, {Rol}, \& {Starling}}]{Pian06}
{Pian}, E., {Mazzali}, P.~A., {Masetti}, N., {et~al.} 2006, \nat, 442, 1011, \dodoi{10.1038/nature05082}

\bibitem[{Pierel {et~al.}(2018)Pierel, Rodney, Avelino, Bianco, Filippenko, Foley, Friedman, Hicken, Hounsell, Jha, Kessler, Kirshner, Mandel, Narayan, Scolnic, \& Strolger}]{pierel_extending_2018}
Pierel, J. D.~R., Rodney, S., Avelino, A., {et~al.} 2018, Publications of the Astronomical Society of the Pacific, 130, 114504, \dodoi{10.1088/1538-3873/aadb7a}

\bibitem[{{Pierel} {et~al.}(2022){Pierel}, {Jones}, {Kenworthy}, {Dai}, {Kessler}, {Ashall}, {Do}, {Peterson}, {Shappee}, {Siebert}, {Barna}, {Brink}, {Burke}, {Calamida}, {Camacho-Neves}, {de Jaeger}, {Filippenko}, {Foley}, {Galbany}, {Fox}, {Gomez}, {Hiramatsu}, {Hounsell}, {Howell}, {Jha}, {Kwok}, {P{\'e}rez-Fournon}, {Poidevin}, {Rest}, {Rubin}, {Scolnic}, {Shirley}, {Strolger}, {Tinyanont}, \& {Wang}}]{Pierel22}
{Pierel}, J.~D.~R., {Jones}, D.~O., {Kenworthy}, W.~D., {et~al.} 2022, \apj, 939, 11, \dodoi{10.3847/1538-4357/ac93f9}

\bibitem[{{Pierel} {et~al.}(2024){Pierel}, {Engesser}, {Coulter}, {Decoursey}, {Siebert}, {Rest}, {Egami}, {Chen}, {Fox}, {Jones}, {Joshi}, {Moriya}, {Zenati}, {Bunker}, {Cargile}, {Curti}, {Eisenstein}, {Gezari}, {Gomez}, {Guolo}, {Johnson}, {Karmen}, {Maiolino}, {Quimby}, {Robertson}, {Shahbandeh}, {Strolger}, {Sun}, {Wang}, \& {Wevers}}]{Pierel24}
{Pierel}, J.~D.~R., {Engesser}, M., {Coulter}, D.~A., {et~al.} 2024, arXiv e-prints, arXiv:2406.05089.
\newblock \doarXiv{2406.05089}

\bibitem[{Pierel {et~al.}(2024)Pierel, Frye, Pascale, Caminha, Chen, Dhawan, Gilman, Grayling, Huber, Kelly, Thorp, Arendse, Birrer, Bronikowski, Canameras, Coe, Cohen, Conselice, Driver, Dsilva, Engesser, Foo, Gall, Garuda, Grillo, Grogin, Henderson, Hjorth, Jansen, Johansson, Kamieneski, Koekemoer, Larison, Marshall, Moustakas, Nonino, Ortiz, Petrushevska, Pirzkal, Robotham, Ryan, Schuldt, Strolger, Summers, Suyu, Treu, Willmer, Windhorst, Yan, Zitrin, Acebron, Chakrabarti, Coulter, Fox, Huang, Jha, Li, Mazzali, Meena, Perez-Fournon, Poidevin, Rest, \& Riess}]{pierel_jwst_2024}
Pierel, J. D.~R., Frye, B.~L., Pascale, M., {et~al.} 2024, arXiv e-prints, arXiv:2403.18954, \dodoi{10.48550/arXiv.2403.18954}

\bibitem[{{Podsiadlowski} {et~al.}(1992){Podsiadlowski}, {Joss}, \& {Hsu}}]{Podsiadlowski+92}
{Podsiadlowski}, P., {Joss}, P.~C., \& {Hsu}, J.~J.~L. 1992, \apj, 391, 246, \dodoi{10.1086/171341}

\bibitem[{{Polin} {et~al.}(2019){Polin}, {Nugent}, \& {Kasen}}]{Polin19}
{Polin}, A., {Nugent}, P., \& {Kasen}, D. 2019, \apj, 873, 84, \dodoi{10.3847/1538-4357/aafb6a}

\bibitem[{{Prentice} {et~al.}(2016){Prentice}, {Mazzali}, {Pian}, {Gal-Yam}, {Kulkarni}, {Rubin}, {Corsi}, {Fremling}, {Sollerman}, {Yaron}, {Arcavi}, {Zheng}, {Kasliwal}, {Filippenko}, {Cenko}, {Cao}, \& {Nugent}}]{Prentice16}
{Prentice}, S.~J., {Mazzali}, P.~A., {Pian}, E., {et~al.} 2016, \mnras, 458, 2973, \dodoi{10.1093/mnras/stw299}

\bibitem[{Rest {et~al.}(2023)Rest, Pierel, Correnti, Canipe, Hilbert, Engesser, Sunnquist, \& Fox}]{rest_arminrestjhat_2023}
Rest, A., Pierel, J., Correnti, M., {et~al.} 2023, arminrest/jhat: {The} {JWST} {HST} {Alignment} {Tool} ({JHAT}),  Zenodo, \dodoi{10.5281/zenodo.7892935}

\bibitem[{{Riess} {et~al.}(2016){Riess}, {Macri}, {Hoffmann}, {Scolnic}, {Casertano}, {Filippenko}, {Tucker}, {Reid}, {Jones}, {Silverman}, {Chornock}, {Challis}, {Yuan}, {Brown}, \& {Foley}}]{Riess16}
{Riess}, A.~G., {Macri}, L.~M., {Hoffmann}, S.~L., {et~al.} 2016, \apj, 826, 56, \dodoi{10.3847/0004-637X/826/1/56}

\bibitem[{{Riess} {et~al.}(2018){Riess}, {Casertano}, {Yuan}, {Macri}, {Bucciarelli}, {Lattanzi}, {MacKenty}, {Bowers}, {Zheng}, {Filippenko}, {Huang}, \& {Anderson}}]{Riess18:gaia}
{Riess}, A.~G., {Casertano}, S., {Yuan}, W., {et~al.} 2018, \apj, 861, 126, \dodoi{10.3847/1538-4357/aac82e}

\bibitem[{{Sahu} {et~al.}(2018){Sahu}, {Anupama}, {Chakradhari}, {Srivastav}, {Tanaka}, {Maeda}, \& {Nomoto}}]{Sahu+18_BL}
{Sahu}, D.~K., {Anupama}, G.~C., {Chakradhari}, N.~K., {et~al.} 2018, \mnras, 475, 2591, \dodoi{10.1093/mnras/stx3212}

\bibitem[{{Sana} {et~al.}(2012){Sana}, {de Mink}, {de Koter}, {Langer}, {Evans}, {Gieles}, {Gosset}, {Izzard}, {Le Bouquin}, \& {Schneider}}]{Sana2012}
{Sana}, H., {de Mink}, S.~E., {de Koter}, A., {et~al.} 2012, Science, 337, 444, \dodoi{10.1126/science.1223344}

\bibitem[{{Sanders} {et~al.}(2012){Sanders}, {Soderberg}, {Levesque}, {Foley}, {Chornock}, {Milisavljevic}, {Margutti}, {Berger}, {Drout}, {Czekala}, \& {Dittmann}}]{Sanders12}
{Sanders}, N.~E., {Soderberg}, A.~M., {Levesque}, E.~M., {et~al.} 2012, \apj, 758, 132, \dodoi{10.1088/0004-637X/758/2/132}

\bibitem[{{Sanders} {et~al.}(2021){Sanders}, {Shapley}, {Jones}, {Reddy}, {Kriek}, {Siana}, {Coil}, {Mobasher}, {Shivaei}, {Dav{\'e}}, {Azadi}, {Price}, {Leung}, {Freeman}, {Fetherolf}, {de Groot}, {Zick}, \& {Barro}}]{Sanders21}
{Sanders}, R.~L., {Shapley}, A.~E., {Jones}, T., {et~al.} 2021, \apj, 914, 19, \dodoi{10.3847/1538-4357/abf4c1}

\bibitem[{{Shahbandeh} {et~al.}(2022){Shahbandeh}, {Hsiao}, {Ashall}, {Teffs}, {Hoeflich}, {Morrell}, {Phillips}, {Anderson}, {Baron}, {Burns}, {Contreras}, {Davis}, {Diamond}, {Folatelli}, {Galbany}, {Gall}, {Hachinger}, {Holmbo}, {Karamehmetoglu}, {Kasliwal}, {Kirshner}, {Krisciunas}, {Kumar}, {Lu}, {Marion}, {Mazzali}, {Piro}, {Sand}, {Stritzinger}, {Suntzeff}, {Taddia}, \& {Uddin}}]{Shahbandeh22}
{Shahbandeh}, M., {Hsiao}, E.~Y., {Ashall}, C., {et~al.} 2022, \apj, 925, 175, \dodoi{10.3847/1538-4357/ac4030}

\bibitem[{{Shen} {et~al.}(2018){Shen}, {Kasen}, {Miles}, \& {Townsley}}]{Shen18a}
{Shen}, K.~J., {Kasen}, D., {Miles}, B.~J., \& {Townsley}, D.~M. 2018, \apj, 854, 52, \dodoi{10.3847/1538-4357/aaa8de}

\bibitem[{{Shivvers} {et~al.}(2017){Shivvers}, {Modjaz}, {Zheng}, {Liu}, {Filippenko}, {Silverman}, {Matheson}, {Pastorello}, {Graur}, {Foley}, {Chornock}, {Smith}, {Leaman}, \& {Benetti}}]{Shivvers17:loss}
{Shivvers}, I., {Modjaz}, M., {Zheng}, W., {et~al.} 2017, \pasp, 129, 054201, \dodoi{10.1088/1538-3873/aa54a6}

\bibitem[{{Smartt} {et~al.}(2015){Smartt}, {Valenti}, {Fraser}, {Inserra}, {Young}, {Sullivan}, {Pastorello}, {Benetti}, {Gal-Yam}, {Knapic}, {Molinaro}, {Smareglia}, {Smith}, {Taubenberger}, {Yaron}, {Anderson}, {Ashall}, {Balland}, {Baltay}, {Barbarino}, {Bauer}, {Baumont}, {Bersier}, {Blagorodnova}, {Bongard}, {Botticella}, {Bufano}, {Bulla}, {Cappellaro}, {Campbell}, {Cellier-Holzem}, {Chen}, {Childress}, {Clocchiatti}, {Contreras}, {Dall'Ora}, {Danziger}, {de Jaeger}, {De Cia}, {Della Valle}, {Dennefeld}, {Elias-Rosa}, {Elman}, {Feindt}, {Fleury}, {Gall}, {Gonzalez-Gaitan}, {Galbany}, {Morales Garoffolo}, {Greggio}, {Guillou}, {Hachinger}, {Hadjiyska}, {Hage}, {Hillebrandt}, {Hodgkin}, {Hsiao}, {James}, {Jerkstrand}, {Kangas}, {Kankare}, {Kotak}, {Kromer}, {Kuncarayakti}, {Leloudas}, {Lundqvist}, {Lyman}, {Hook}, {Maguire}, {Manulis}, {Margheim}, {Mattila}, {Maund}, {Mazzali}, {McCrum}, {McKinnon}, {Moreno-Raya}, {Nicholl}, {Nugent}, {Pain}, {Pignata}, {Phillips}, {Polshaw}, {Pumo}, {Rabinowitz},
  {Reilly}, {Romero-Ca{\~n}izales}, {Scalzo}, {Schmidt}, {Schulze}, {Sim}, {Sollerman}, {Taddia}, {Tartaglia}, {Terreran}, {Tomasella}, {Turatto}, {Walker}, {Walton}, {Wyrzykowski}, {Yuan}, \& {Zampieri}}]{Smartt15_pessto}
{Smartt}, S.~J., {Valenti}, S., {Fraser}, M., {et~al.} 2015, \aap, 579, A40, \dodoi{10.1051/0004-6361/201425237}

\bibitem[{{Smith} {et~al.}(2015){Smith}, {Mauerhan}, {Cenko}, {Kasliwal}, {Silverman}, {Filippenko}, {Gal-Yam}, {Clubb}, {Graham}, {Leonard}, {Horst}, {Williams}, {Andrews}, {Kulkarni}, {Nugent}, {Sullivan}, {Maguire}, {Xu}, \& {Ben-Ami}}]{Smith15}
{Smith}, N., {Mauerhan}, J.~C., {Cenko}, S.~B., {et~al.} 2015, \mnras, 449, 1876, \dodoi{10.1093/mnras/stv354}

\bibitem[{{Stahl} {et~al.}(2020){Stahl}, {Zheng}, {de Jaeger}, {Brink}, {Filippenko}, {Silverman}, {Cenko}, {Clubb}, {Graham}, {Halevi}, {Kelly}, {Kleiser}, {Shivvers}, {Yuk}, {Cobb}, {Fox}, {Kandrashoff}, {Kong}, {Mauerhan}, {Wang}, \& {Wang}}]{Stahl20}
{Stahl}, B.~E., {Zheng}, W., {de Jaeger}, T., {et~al.} 2020, \mnras, 492, 4325, \dodoi{10.1093/mnras/staa102}

\bibitem[{{Sun} {et~al.}(2022){Sun}, {Maund}, {Crowther}, {Hirai}, {Kashapov}, {Liu}, {Liu}, \& {Zapartas}}]{sun21}
{Sun}, N.-C., {Maund}, J.~R., {Crowther}, P.~A., {et~al.} 2022, \mnras, 510, 3701, \dodoi{10.1093/mnras/stab3768}

\bibitem[{{Taddia} {et~al.}(2015){Taddia}, {Sollerman}, {Leloudas}, {Stritzinger}, {Valenti}, {Galbany}, {Kessler}, {Schneider}, \& {Wheeler}}]{Taddia15}
{Taddia}, F., {Sollerman}, J., {Leloudas}, G., {et~al.} 2015, \aap, 574, A60, \dodoi{10.1051/0004-6361/201423915}

\bibitem[{{Taddia} {et~al.}(2016){Taddia}, {Moquist}, {Sollerman}, {Rubin}, {Leloudas}, {Gal-Yam}, {Arcavi}, {Cao}, {Filippenko}, {Graham}, {Mazzali}, {Nugent}, {Pan}, {Silverman}, {Xu}, \& {Yaron}}]{Taddia16}
{Taddia}, F., {Moquist}, P., {Sollerman}, J., {et~al.} 2016, \aap, 587, L7, \dodoi{10.1051/0004-6361/201527983}

\bibitem[{{Taddia} {et~al.}(2019){Taddia}, {Sollerman}, {Fremling}, {Barbarino}, {Karamehmetoglu}, {Arcavi}, {Cenko}, {Filippenko}, {Gal-Yam}, {Hiramatsu}, {Hosseinzadeh}, {Howell}, {Kulkarni}, {Laher}, {Lunnan}, {Masci}, {Nugent}, {Nyholm}, {Perley}, {Quimby}, \& {Silverman}}]{Taddia19}
{Taddia}, F., {Sollerman}, J., {Fremling}, C., {et~al.} 2019, \aap, 621, A71, \dodoi{10.1051/0004-6361/201834429}

\bibitem[{{Tucker} {et~al.}(2024){Tucker}, {Hinkle}, {Angus}, {Auchettl}, {Hoogendam}, {Shappee}, {Kochanek}, {Ashall}, {de Boer}, {Chambers}, {Desai}, {Do}, {Fulton}, {Gao}, {Herman}, {Huber}, {Lidman}, {Lin}, {Lowe}, {Magnier}, {Martin}, {Minguez}, {Nicholl}, {Pursiainen}, {Smartt}, {Smith}, {Srivastav}, {Tucker}, \& {Wainscoat}}]{Tucker24}
{Tucker}, M.~A., {Hinkle}, J., {Angus}, C.~R., {et~al.} 2024, arXiv e-prints, arXiv:2405.00113, \dodoi{10.48550/arXiv.2405.00113}

\bibitem[{{Valenti} {et~al.}(2016){Valenti}, {Howell}, {Stritzinger}, {Graham}, {Hosseinzadeh}, {Arcavi}, {Bildsten}, {Jerkstrand}, {McCully}, {Pastorello}, {Piro}, {Sand}, {Smartt}, {Terreran}, {Baltay}, {Benetti}, {Brown}, {Filippenko}, {Fraser}, {Rabinowitz}, {Sullivan}, \& {Yuan}}]{valenti16}
{Valenti}, S., {Howell}, D.~A., {Stritzinger}, M.~D., {et~al.} 2016, \mnras, 459, 3939, \dodoi{10.1093/mnras/stw870}

\bibitem[{{Van Dyk} {et~al.}(2016){Van Dyk}, {de Mink}, \& {Zapartas}}]{vandyk16}
{Van Dyk}, S.~D., {de Mink}, S.~E., \& {Zapartas}, E. 2016, \apj, 818, 75, \dodoi{10.3847/0004-637X/818/1/75}

\bibitem[{{Woosley}(1993)}]{Woosley93}
{Woosley}, S.~E. 1993, \apj, 405, 273, \dodoi{10.1086/172359}

\bibitem[{{Woosley} \& {Bloom}(2006{\natexlab{a}})}]{Woosley&Bloom06}
{Woosley}, S.~E., \& {Bloom}, J.~S. 2006{\natexlab{a}}, \araa, 44, 507, \dodoi{10.1146/annurev.astro.43.072103.150558}

\bibitem[{{Woosley} \& {Bloom}(2006{\natexlab{b}})}]{Woosley06}
---. 2006{\natexlab{b}}, \araa, 44, 507, \dodoi{10.1146/annurev.astro.43.072103.150558PDF: http://arjournals.annualreviews.org/doi/pdf/10.1146/annurev.astro.43.072103.150558}

\bibitem[{{Woosley} {et~al.}(1994){Woosley}, {Eastman}, {Weaver}, \& {Pinto}}]{Woosley+94_IIb}
{Woosley}, S.~E., {Eastman}, R.~G., {Weaver}, T.~A., \& {Pinto}, P.~A. 1994, \apj, 429, 300, \dodoi{10.1086/174319}

\bibitem[{{Woosley} \& {MacFadyen}(1999)}]{Woosley99}
{Woosley}, S.~E., \& {MacFadyen}, A.~I. 1999, \aaps, 138, 499, \dodoi{10.1051/aas:1999325}

\bibitem[{{Woosley} \& {Weaver}(1995)}]{WoosleyWeaver95}
{Woosley}, S.~E., \& {Weaver}, T.~A. 1995, \apjs, 101, 181, \dodoi{10.1086/192237}

\bibitem[{{Yoon} {et~al.}(2010){Yoon}, {Woosley}, \& {Langer}}]{Yoon&Woosley10}
{Yoon}, S.~C., {Woosley}, S.~E., \& {Langer}, N. 2010, \apj, 725, 940, \dodoi{10.1088/0004-637X/725/1/940}

\bibitem[{{Zapartas} {et~al.}(2017){Zapartas}, {de Mink}, {Van Dyk}, {Fox}, {Smith}, {Bostroem}, {de Koter}, {Filippenko}, {Izzard}, {Kelly}, {Neijssel}, {Renzo}, \& {Ryder}}]{zapartas17}
{Zapartas}, E., {de Mink}, S.~E., {Van Dyk}, S.~D., {et~al.} 2017, \apj, 842, 125, \dodoi{10.3847/1538-4357/aa7467}

\bibitem[{{Zenati} {et~al.}(2020){Zenati}, {Siegel}, {Metzger}, \& {Perets}}]{Zenati+20}
{Zenati}, Y., {Siegel}, D.~M., {Metzger}, B.~D., \& {Perets}, H.~B. 2020, \mnras, 499, 4097, \dodoi{10.1093/mnras/staa3002}

\end{thebibliography}
\bibliographystyle{aasjournal}

\end{document}